\documentclass[12pt]{article}
\usepackage{amssymb,amsmath}
\usepackage{bm} 
\usepackage{booktabs} 
\usepackage{array}
\usepackage{latexsym}
\usepackage{graphicx}
\usepackage{color}
\usepackage{slashed}
\usepackage{datetime}
\usepackage[nosort]{cite}
\usepackage{verbatim}
\usepackage{enumerate}
\usepackage{makecell} 
\usepackage{chngpage} 
\allowdisplaybreaks
\usepackage{psfrag}
\usepackage{mathtools}
\usepackage{mciteplus}
\usepackage{dsfont}
\usepackage{multirow} 
\usepackage{arydshln} 

\usepackage[colorlinks=true,      linkcolor=blue,      urlcolor=blue,      
filecolor=blue,      citecolor=blue,       pdfstartview=FitH,     
pdfpagemode=UseNone,      bookmarksopen=true]{hyperref}  
\usepackage[all]{hypcap}     

\newcommand{\comm}[1]{} 

\newcommand{\coo}[1]{{\mathcal{O}\left(#1\right)}}

\newcommand{\kett}[1]{{\left|{#1}\right\rangle}}

\def\({\left(}
\def\){\right)}
\def\[{\left[}
\def\]{\right]}

\def\pd{{\partial}}

\def\One{{\hbox{ 1\kern-.8mm l}}}

\def\barray{\begin{array}}
	\def\earray{\end{array}}
\def\be{\begin{equation}}
	\def\ee{\end{equation}}
\def\bea{\begin{eqnarray}}
	\def\eea{\end{eqnarray}}
\def\bal{\begin{align}}
	\def\eal{\end{align}}

\def\bt{{\widetilde{b}}}

\def\Gt{{\widetilde{G}}}

\def\Jt{{\widetilde{J}}}

\def\Lt{{\widetilde{L}}}

\def\phit{{\widetilde{\phi}}}
\def\psit{{\widetilde{\psi}}}

\def\Psit{{\widetilde{\Psi}}}

\newcommand{\oO}{\overline{\Omega}}

\newcommand{\hb}{{\overline{h}}}
\newcommand{\jb}{{\overline{j}}}
\newcommand{\hbn}{{\overline{h}^{\mathrm{ NS}}}}
\newcommand{\jbn}{{\overline{j}^{\mathrm{ NS}}}}

\numberwithin{equation}{section} 
\allowdisplaybreaks  

\makeatletter
\g@addto@macro\bfseries{\boldmath}
\makeatother

\definecolor{cardinal}{rgb}{0.6,0,0}
\definecolor{darkgreen}{rgb}{0,0.4,0}
\definecolor{golden}{rgb}{0.92, 0.7, 0}
\definecolor{midnight}{rgb}{0, 0, 0.5}
\definecolor{darkblue}{rgb}{0, 0, 0.7}
\definecolor{purple}{rgb}{0.5, 0, 0.5}

\def\IR{\mathbb{R}}

\def\cD{{\cal D}}
\def\cF{{\cal F}}

\def\cL{{\cal L}}
\def\cM{{\cal M}}

\def\cft{{\cal F}}

\def\cO{{\cal O}}

\def\Jt{\tilde{J}}

\def\vh{\hat{v}}

\def\cLh{\widehat{\mathcal{L}}}

\newcommand{\Jb}{\bar{J}}
\newcommand{\Lb}{\bar{L}}

\def\hb{{\bar{h}}}

\def\tF{{\tilde F}}
\def\tA{{\tilde A}}

\def\Psit{{\widetilde \Psi}}

\usepackage{epsf}
\usepackage{cite}
\usepackage{fancyhdr}
\usepackage{bm}
\usepackage{enumerate} 
	
	\usepackage{empheq} \empheqset{box=\fbox}
	\usepackage[usenames,dvipsnames]{xcolor}
	
	\numberwithin{equation}{section}  
	\allowdisplaybreaks

	\usepackage{graphicx}
	\usepackage{tikz}
	\usetikzlibrary{decorations.markings}
	\tikzset{->-/.style={decoration={
				markings,
				mark=at position #1 with {\arrow{stealth}}},postaction={decorate}}}
	\usepackage{adjustbox}
	\usepackage{pgfplots}
	\pgfplotsset{compat=1.11}
	\usepgfplotslibrary{fillbetween}
	\usetikzlibrary{intersections}
	\usetikzlibrary{patterns}
	
	\pgfdeclarelayer{bg}
	\pgfsetlayers{bg,main}
	
	\usetikzlibrary{calc}
	\tikzset{
		samples=100,
	}
	\pgfplotsset{compat=1.11}
	
	\pgfmathsetmacro\T{3.14}
	\pgfmathsetmacro\A{0.2}
	\pgfmathsetmacro\N{4}
	\pgfmathsetmacro\D{\N*\T}
	
\begin{document}
		
		
		\begin{flushright}
			
		\end{flushright}
		
		\vspace{3mm}
		
		\begin{center}

			{\huge {\bf Vector Superstrata: Part Two}}
			
			\vspace{14mm}
			
			{\large
				\textsc{
    \vspace{14mm}
					Nejc \v{C}eplak$^1$ and  Shaun D. Hampton$^2$}} 
\textit{
     \centerline{$^1$ School of Mathematics and Hamilton Mathematics Institute,}
    \centerline{Trinity College,}
    \centerline{Dublin 2, Ireland}}

    \textit{
     \centerline{$^2$ School of Physics}
\centerline{Korea Institute for Advanced Study,}
\centerline{85 Hoegiro Dongdaemun-gu,}
\centerline{Seoul, 02455, Korea.}
 }

\medskip

\vspace{4mm} 
%

{\footnotesize\upshape\ttfamily  ceplakn @ tcd.ie, sdh2023 @ kias.re.kr
 } \\
\vspace{13mm}

\textsc{Abstract}

\end{center}

\begin{adjustwidth}{10mm}{10mm} 

\vspace{1mm}
\noindent

Microstate geometries are proposed microstates of black holes which can be described within supergravity.
Even though their number may not reproduce the full entropy of black holes with finite-sized horizons, they still offer a glimpse into the microscopic structure of black holes. 
In this paper we construct a new set of microstate geometries of the supersymmetric D1-D5-P black hole, where the momentum charge is carried by a vector field, as seen from the perspective of six-dimensional supergravity. 
To aid our construction, we develop an algorithm which solves a complicated partial differential equation using the regularity of the geometries.
The new solutions are asymptotically AdS$_3\times S^3$, and have a long, but finite AdS$_2$ throat that caps off without ever developing a horizon. 
These microstate geometries have a holographic interpretation as coherent superpositions of heavy states in the boundary D1-D5 CFT.
We identify the states which are dual to our newly constructed solutions and carry out some basic consistency checks to support our identification.

\end{adjustwidth}

\thispagestyle{empty}
 \setcounter{page}{0}
\clearpage



\baselineskip=14pt
\parskip=3pt

\tableofcontents

\baselineskip=18pt
\parskip=3pt



\thispagestyle{empty}
\clearpage
\addtocounter{page}{-1}

\clearpage

\section{Introduction}
\label{sec:Intro}

The existence of black holes perfectly embodies and necessitates the synchronicity between quantum mechanics and general relativity.
In particular, the non-vanishing black-hole entropy indicates that there may exist a more fundamental set of microstates accounting for this entropy. 
However, the explicit form of these microstates may be hard to realize if one assumes only the degrees of freedom found in general relativity.

A framework which offers additional degrees of freedom is string theory. 
%
%
In particular, the fuzzball paradigm \cite{Mathur:2005zp,Bena:2022rna,Bena:2022ldq} states that a classical black hole
should be replaced by objects in string theory, called fuzzballs%
, which are horizonless and, when averaged over, reproduce the same observable effects as a black hole of the same mass, angular momentum, and charges.%
    \footnote{However, for extremal black holes with vanishing horizon radius the corresponding averaged configurations are larger than the stretched horizon \cite{Martinec:2023xvf, Martinec:2023gte}.}
In general, fuzzballs are highly quantum and stringy configurations.
However, analogous to how a system of atoms, through processes of stimulated emission, can be configured to produce a coherent monochromatic beam of photons, a laser,  one can also consider very coherent fuzzball states which admit a smooth, geometric description. 
This subset of fuzzballs, often called microstate geometries \cite{Bena:2013dka, Bena:2022sge}, can be described using just supergravity,  the low-energy limit of string theory, and  is the focus of this paper.

We work in the D1-D5 system:
Starting with Type IIB string theory on%
\footnote{$\mathcal{M}_4$ can be either $T^4$ or $K3$. In this paper we use exclusively $T^4$.}
$\mathbb{R}^{4,1}\times S_y^1\times \mathcal{M}_4$ 
one wraps $N_1$ D1-branes along the $y$-circle and $N_5$ D5-branes along $S_y^1\times \mathcal{M}_4$.
Such two-charge configurations preserve 8 out of the total 32 supercharges and we  refer to them as 1/4-BPS. 
By adding momentum charge, $P$, along $S_y^1$, one breaks  4 additional supercharges leading to 1/8-BPS configurations.
The degeneracy of bound states with such charges, as calculated at weak string coupling, matches the (exponent of) the Bekenstein-Hawking entropy of the corresponding D1-D5-P black hole \cite{Strominger:1996sh, Callan:1996dv}. 
While this provides a match between the ensembles of black hole microstates and ensembles of brane configurations as the string coupling (or the gravitational coupling, $G_N$) is varied, we are interested in the fate of individual microstates. 

The first $1/4$-BPS  microstates were constructed in \cite{Lunin:2001jy,  Lunin:2001fv, Lunin:2002iz} by wrapping fundamental strings along  $S_y^1$ and allowing for transverse oscillations  with which the string carries momentum.
These configurations can then be, through a series of S and T-dualities, related to the D1-D5 system \cite{Taylor:2005db,Kanitscheider:2006zf, Kanitscheider:2007wq, Skenderis:2008qn}. 
In the latter frame, the near-horizon region of (geometric) microstates contains an AdS$_3$ factor, which, through the AdS/CFT correspondence \cite{Maldacena:1997re}, allows us to relate the geometries to Ramond-Ramond (RR) ground states in the dual two-dimensional D1-D5 CFT \cite{ Lunin:2001jy, Lunin:2001fv,Taylor:2005db, Kanitscheider:2006zf, Kanitscheider:2007wq}.

 The maximally spinning round supertube, perhaps the simplest of such solutions, is  the starting point of our construction. 
By adding a small perturbation on top of this solution one can generate the additional momentum charge. 
Utilising the linear and upper-triangular structure of the relevant BPS equations \cite{Bena:2011dd, Giusto:2013rxa, Cano:2018wnq, Ceplak:2022wri}, one is then able to find the fully backreacted, $1/8$-BPS microstate geometries, called superstrata \cite{Bena:2015bea, Bena:2016agb,Bena:2016ypk, Bena:2017xbt, Ceplak:2018pws, Heidmann:2019zws, Heidmann:2019xrd, Shigemori:2020yuo}.
When viewed as solutions of six-dimensional supergravity, in most superstrata the momentum is carried by a field in the tensor multiplet.
However, when counted, such solutions represent a parametrically small contribution to the entropy of the D1-D5-P black hole \cite{Shigemori:2019orj, Mayerson:2020acj}.

Recently, it has been shown that the momentum charge can also be carried by vector fields \cite{Bena:2022sge} with the corresponding set of microstate geometries being dubbed ``vector superstrata'' \cite{Ceplak:2022pep}.
In this paper, we construct a new family of such vector superstrata.
Our work aims to complete the set of $1/8$-BPS microstate geometries of the D1-D5-P black hole which can be constructed in non-orbifolded AdS$_3$ \cite{Bena:2016agb,Shigemori:2022gxf}.
We identify their dual CFT states and perform some rudimentary checks that support our identification. 
Furthermore, we develop a systematic method to solve the BPS equations and  present an algorithmic way to solve a  Laplace equation in spheroidal coordinates with arbitrarily complicated sources.
We do so by making an informed ansatz  based on smoothness of the geometries in the interior and their prescribed asymptotic behaviour.
We use an appropriate basis expansion, which reduces the partial differential equation to a system of algebraic equations for the basis coefficients.
%
We hope that the techniques presented will prove helpful for future constructions of microstate geometries and related computations.

In Section~\ref{sec:CFT} we describe CFT states which are dual to the vector superstrata  that we construct in this paper.
In Section~\ref{sec:BPSeq} we briefly review the relevant supersymmetric ansatz and the associated BPS equations that are solved by the new microstate geometries.
In Section~\ref{sec:Gravity} we start with a vector perturbation and show how to find a fully backreacted solution.  In Section~\ref{sec:Bootstrapy} we  employ a type of `bootstrap' technique to find the explicit solutions. In the final section we discuss our results and conclude with  some  future directions. 
The appendices contain details omitted in the main body of the paper.
In Appendix~\ref{app:CFT} we summarise the conventions used in the CFT picture. 
Appendix~\ref{app:SolLay} contains the detailed calculations that explicitly show how the constructed geometries solve the aforementioned BPS equations.
Finally, in Appendix~\ref{app:example}, we present in full detail a particular solution as an application of our newly developed method to solve the full system of BPS equations.

\section{Motivation from the CFT picture}
\label{sec:CFT}

The bulk of this paper deals with solving BPS equations of six-dimensional supergravity in order to obtain regular solutions that asymptote to AdS$_3\times S^3$. 
Through the AdS/CFT correspondence \cite{Maldacena:1997re}, such solutions have an equivalent description in terms of heavy states in the dual D1-D5 CFT. This is a two-dimensional supersymmetric CFT whose central charge is $c = \bar c = 6\,N$, where $N = N_1\,N_5$ is the product of the integer number of D1-branes and D5-branes used to construct the system.
In this section, we present states that are dual to the geometries that we construct in subsequent sections.
We limit ourselves only to the information that is immediate to our construction, purposefully omitting a lot of details (or relegating them to Appendix~\ref{app:CFT}, where we also summarise our conventions).%
\footnote{A more thorough analysis can be found for example in \cite{Kanitscheider:2006zf, Kanitscheider:2007wq, Skenderis:2008qn, Avery:2010qw,Shigemori:2019orj, Shigemori:2020yuo}, while states related to microstate geometries with vector fields were discussed in detail in \cite{Ceplak:2022pep}.}

To describe the geometries in the CFT language, we use the free orbifold point. 
In the bulk this free CFT corresponds to strings on AdS$_3$ in the tensionless limit \cite{Eberhardt:2018ouy, Eberhardt:2019ywk, Eberhardt:2020akk}, 
while the dual of the supergravity theory is a strongly coupled CFT, over which we have little control. 
However, since all of our states preserve at least 4 supercharges,  we rely on non-renormalisation theorems \cite{Baggio:2012rr} to protect some properties of supersymmetric states -- in particular, the charges under the symmetry group  --   from changing as we move in the moduli space of the theory. 

\begin{figure}[t]
    \centering
    \includegraphics[width =0.9\textwidth]{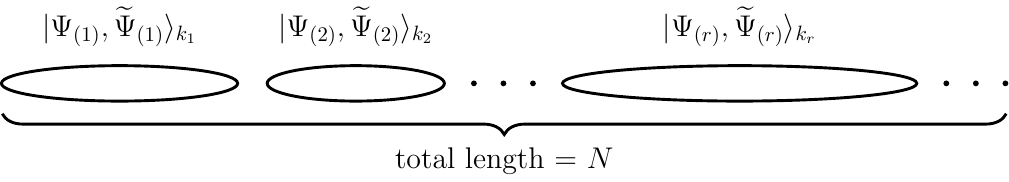}
    \caption{A schematic depiction of a state at the orbifold point CFT: On each strand of length $k_r$ we specify a state $|\Psi_{(r)}, \Psit_{(r)}\rangle_{k_r}$. The total state is  the product of states on all strands \eqref{eq:GeneralState}, subject to  the sum of all strand lengths being $N$ and the full state being invariant under the permutation group $S_N$.}
    \label{fig:State1}
\end{figure}

The target space of the theory with $\mathcal{M}_4 = T^4$ at the symmetric orbifold point is $(T^4)^N/S_N$ --  $N$ copies of a free, $c=\bar c=6$ CFT with target space $T^4$ identified under the permutation group $S_N$. 
Each individual copy can be thought of as living on an effective closed string, which we refer to as a strand, described by $t$ and $y$ coordinates that parameterise the boundary of AdS$_3$. 
The theory, including its supersymmetry group, $SU(1,1|2)_L \times SU(1,1|2)_R$, naturally splits into left-moving and right-moving sectors. 
The state on each strand, labelled by $(r)$, is then written as $|\Psi_{(r)},\Psit_{(r)}\rangle_{k_r}$, where  $\Psi$ and $\Psit$ are the states in the left-moving and right-moving sectors, respectively. 
The integer $k_r$ denotes the length of the effective string: Namely,  one is able to make an effectively longer strand by changing the boundary conditions on the periodic coordinate $y$.
The strand length can be arbitrary as long as the total length of all strands combined is equal to $N$.
The full state is then the permutation invariant product of states in all copies $(r)$, subject to this constraint (for a schematic representation see Figure~\ref{fig:State1}).

In what follows, the invariance under permutations does not play a critical role, so we suppress the copy index $(r)$. 
We can thus write the full state as a product over all possible individual states on all possible strand lengths, which we can schematically write as
\begin{align}
\label{eq:GeneralState}
   \prod_{\Psi, \Psit}\,\prod_{k=1}^{N} \left(|\Psi,\Psit\rangle_k\right)^{N_k^{\Psi,\Psit}}\,,
\end{align}
where $N_k^{\Psi,\Psit}$ denotes the number of states $|\Psi,\Psit\rangle_k$.
The state \eqref{eq:GeneralState} is constrained by 
\begin{align}
\label{eq:StrandBudget}
   \sum_{\Psi, \Psit}\, \sum_{k=1}^{N}k\,N_k^{\Psi,\Psit} = N\,,
\end{align}
which ensures that the total strand length is $N$. 

We are interested in supersymmetric heavy states in this theory -- states which preserve a finite number of supercharges and whose conformal dimensions scale with the total central charge, $\Delta\sim c\sim N$.
In the main text we work primarily in the Ramond  sector of the theory, since states in this sector arise naturally when one considers the near horizon region of asymptotically flat supersymmetric black holes \cite{Maldacena:1998bw}.  
In particular, the 1/4-BPS Ramond ground states have a large degeneracy and their counting reproduces the entropy of the two-charge D1-D5 black hole \cite{Strominger:1996sh, Sen:1995in, Lunin:2001jy}. 
These ground states are the starting point of our construction. 

On each strand, one can have 16 different states, 8 of them are bosonic while the remaining 8 are fermionic. 
Let us begin by focusing on the set of four bosonic states which, following \cite{Shigemori:2019orj, Shigemori:2020yuo}, can be labelled by their eigenvalues under the $R$-symmetry group $SU(2)_L\times SU(2)_R$%
\footnote{The $R$-symmetry $SU(2)_L\times SU(2)_R\simeq SO(4)$ is associated with the isometries of the $S^3$ in the bulk. Intuitively one can think of labelling states with the $j$ and $\jb$ eigenvalues as labelling them according to their angular momentum in two orthogonal planes of rotation in $S^3$.}
\begin{align}
\label{eq:Groundstate1}
    \kett{\alpha =\pm, \dot \alpha =\pm }_k\,, \qquad j = \frac{\alpha}2\,,\quad \jb = \frac{\dot \alpha}{2}\,,
\end{align}
with the eigenvalues under the $SL(2,\mathbb{R})_L\times SL(2,\mathbb{R})_R$ symmetry being $h = \hb = k/4$.%
\footnote{The eigenvalues under the conformal symmetries are connected to the isometries of the AdS$_3$ space and are related to the mass and linear momentum in this spacetime. In particular $\Delta = h+ \hb$ and $n_P = h - \hb$.}
Coherent superpositions of states containing only these strands can be realised in the bulk as  Lunin-Mathur geometries \cite{Lunin:2001jy, Lunin:2002iz, Taylor:2005db, Kanitscheider:2006zf, Kanitscheider:2007wq, Bena:2015bea,Bena:2017xbt}. 
Perhaps the simplest geometry is the maximally spinning supertube, which is dual to a state consisting only of $\kett{++}_1$ strands
\begin{align}
    \label{eq:Supertubestate}
    \kett{\text{Supertube}} = \left(\kett{++}_1\right)^N\,, \qquad j = \jb = \frac{N}{2}\,,\qquad h = \hb = \frac{N}{4} = \frac{c}{24} \,,
\end{align}
with the conformal dimension, $\Delta = h + \hb$, indeed scaling with the central charge. 

In this paper we construct new 1/8-BPS solutions via the standard six-dimensional superstratum method \cite{Bena:2015bea, Bena:2016agb, Bena:2016ypk, Bena:2017xbt, Ceplak:2018pws, Heidmann:2019zws, Heidmann:2019xrd, Shigemori:2020yuo,  Ceplak:2022pep}.
One begins by adding a small perturbation to the supertube geometry -- in the CFT this corresponds to adding to the supertube state \eqref{eq:Supertubestate} a small amount of strands dual to the perturbation in the bulk. 
Then one acts on this new state with the generators $L_n$, $J_n^i$, $G_n^{\alpha A}$ of the left-moving supersymmetry group,  $SU(1,1|2)_L$.
This increases the momentum charge of the state, $n_P = h - \hb$, while preserving some supersymmetry due the right-moving sector remaining unchanged.
Finally, to obtain the fully backreacted solution in the bulk, one needs to solve the full equations of supergravity.
On the CFT side, this corresponds to increasing the number of strands associated with the perturbation and taking a coherent superposition of such states -- this ensures that the state has a valid description within supergravity.

The deformation that is central for this paper is described by the state%
\footnote{It is natural to arrive at this state by starting in the Neveu-Schwarz sector and then spectral flowing to the Ramond sector. We present this path  in Appendix~\ref{app:CFT}.}
\begin{align}
	\label{eq:Rstate}
	\kett{k,m,n;  A,\dot A} \equiv \left(L_{-1}-J_{-1}^3\right)^{n}\, \left(J_{-1}^+\right)^{m-1} \, G_{-1}^{+A}\,\kett{+\dot A}_k\,,
\end{align}
whose quantum numbers are given by
\begin{align}
	h = \frac{k}{4} + m + n\,, \qquad j = m\,, \qquad \hb = \frac{k}4\,,\qquad  \jb = 0\,,
\end{align}
and $k = 2, 3, \ldots$, $m = 1,2, \ldots k-1$, and $n = 0,1,2\ldots$.
This state is based on a fermionic 1/4-BPS state $|+\dot A\rangle_k$ on a strand of length $k$. The index $\dot A=1,2$ denotes a doublet under the $SU(2)_C$, which is part of the symmetry related to the isometry of the $T^4$, $SU(2)_B\times SU(2)_C \simeq SO(4)_I$.
The residual $SU(2)_B$ appears through the index $A=1,2$ of the supersymmetry generator $G_{-1}^{+A}$. The action of this generator is two-fold:
It transforms the fermonic perturbation into a bosonic one and the two indices $A$ and $\dot A$ combine into a vector index of $SO(4)_I$.
As a consequence, the gravity dual of \eqref{eq:Rstate} is described by a vector field when viewed in six-dimensional supergravity compactified on $T^4$.

\begin{figure}[t]
    \centering
    \includegraphics[width=\textwidth]{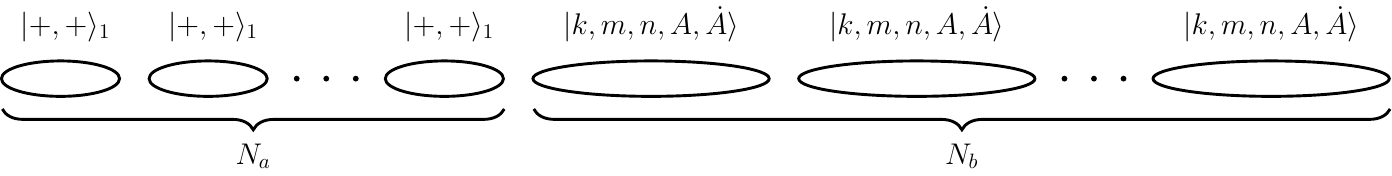}
    \caption{Schematic depiction of the state whose bulk dual we construct in later sections: We combine $N_a$ copies of states which appear in the round supertube and $N_b$ copies of states defined in \eqref{eq:Rstate}. }
    \label{fig:stares2}
\end{figure}

The full state that we consider is then given by a coherent sum (see also Figure~\ref{fig:stares2})
\begin{align}
	\label{eq:FullStateR}
	\kett{\psi} \sim \sum_{N_a, N_b}{}^{\!\!\!'}\left(\kett{++}_1\right)^{N_a}\, \left(\kett{k,m,n;  A, \dot A} \right)^{N_b}\,,
\end{align}
where 
\begin{align}
    \label{eq:ActualConstraint}
    N=N_a + k\,N_b\,,
\end{align}
and $\sum{}^{\!'}$ denotes the coherent sum.
As discussed in the introduction, generic distributions of $N_a$ and $N_b$ describe a highly quantum state which cannot be reliably described within supergravity. Taking a coherent superposition of such quantum states is necessary for the dual of \eqref{eq:FullStateR} to have a valid description in terms of a geometric solution \cite{Kanitscheider:2006zf, Skenderis:2006ah, Kanitscheider:2007wq, Giusto:2015dfa, Bena:2017xbt}.
However, since the supergravity description is valid in the limit where $N_a, N_b \gg 1$ the coherent sum is sharply peaked around its average values.
To a good approximation we can then use these average values to describe the states \eqref{eq:FullStateR} and its dual geometry.
To lighten the notation, we still use $N_a$ and $N_b$ to denote these averages. 
In terms of these, the total conserved charges of the state \eqref{eq:FullStateR} are
\begin{align}
	\label{eq:FullRCharges}
	h =  \frac{N}4 + N_b \left(m+n\right)\,, \qquad \hb = \frac{N}4\,, \qquad j = \frac{N_a}{2} + N_b\, m\,, \qquad \jb = \frac{N_a}{2}\,.
\end{align}
Since, $h> \hb$, such states carry a non-trivial momentum charge
\begin{align}
	\label{eq:MomChargeCFT}
	n_P = h- \hb =  N_b \left(m+n\right)\,,
\end{align} 
which should also be seen from the dual gravity description. Indeed, this will be a sanity check of our fully backreacted solutions.

\section{Supersymmetric ansatz and BPS equations}
\label{sec:BPSeq}

Coherent superpositions of heavy states in the D1-D5 CFT have a dual representation in terms of supergravity solutions that are asymptotically AdS$_3 \times S^3\times T^4$.
While in general this requires us to work with ten-dimensional supergravity, one usually assumes that the geometries do not depend on the coordinates of the four-torus, in which case the system reduces to six-dimensional supergravity. 

The minimal field content that needs to be included in the lower-dimensional theory depends on the states one wants to describe. 
For the state \eqref{eq:FullStateR}, it is sufficient to look at minimal supergravity coupled to a tensor and a vector multiplet.
The bosonic fields contained in such a theory are the metric $g_{MN}$, a two-form gauge field $B_{MN}$, a one-form vector field $A_M$, and a dilaton $\phi$. 

Supersymmetric solutions of such theories are completely characterised \cite{Gutowski:2003rg,Cariglia:2004kk, Cano:2018wnq}. 
The six-dimensional metric in the Einstein frame can be decomposed as
\begin{align}
		ds_6^2  ~&=~-\frac{2}{\sqrt{Z_1\, Z_2}}(d v+\beta)\Big[d u+\omega + \frac{\cF}{2}(d v+\beta)\Big]~+~\sqrt{Z_1\, Z_2}\,ds_4^2\,,	\label{eq:MetAns1}
\end{align}
where $u$ and $v$ are null-coordinates that are related to the usual time, $t$, and the coordinate parametrising the asymptotic circle of AdS$_3$, $y\sim y + 2\pi\,R_y$, through
\begin{align}\label{eq:DefNullCoord}
	v \equiv \frac{t + y}{\sqrt 2}\,,\qquad u \equiv \frac{t-y}{\sqrt 2}\,.
\end{align}
The  $ds_4^2$ denotes the line-element of a four-dimensional Euclidean space, which we  refer to as the \emph{base space}.
%
The remaining fields in the theory can be written as
\begin{subequations}
	\label{eq:Potentials}
	\begin{align}
		A ~& =~ \frac{Z_A}{Z_2}\,(dv + \beta) - \tA\,, \qquad e^{2\phi} = \frac{Z_2}{Z_1}\\*
		B ~& =~ -\frac{1}{Z_2}\, (du +\omega)\wedge (dv + \beta)+ a_2 \wedge (dv + \beta) + \gamma_1\,.
	\end{align}
\end{subequations}
where $Z_1$, $Z_2$, $Z_A$, and $\cF$ are scalar functions, while $\beta$, $\omega$, $\tA$ and $a_2$ are one-forms and $\gamma_1$ is a two-form on the base space.
Supersymmetry also demands that (the components of) these ansatz quantities are independent of one null coordinate, which we choose to be $u$, so that the ansatz quantities can be functions of $v$ and the base space coordinates only.

The power of this decomposition is that one can recast the BPS equations as differential equations for the ansatz quantities on the four-dimensional base space. 
These equations can then be organised in several layers with an upper-triangular structure: The solutions to the previous layer act as (at most) quadratic sources for the next layer \cite{Bena:2011dd, Giusto:2013rxa,Cano:2018wnq, Ceplak:2022wri}. 
Anticipating the form of the solutions, we now present a simplified version of the BPS equations that are relevant in the construction that follows. 

The initial BPS equations that one needs to solve are in the so-called ``zeroth layer'', which determine the base space and the one-form $\beta$.
We take the base space to be given by flat $\mathbb{R}^4$ and the one-form to be $v$-independent,  $\dot \beta =0$.
This automatically satisfies all the equations in the zeroth layer, apart from the self-duality condition
\begin{align}
    \label{eq:BetaSD}
    *_4d_4\beta ~=~ d_4\beta\,,
\end{align}
where  $d_4$ is the exterior derivative restricted to the base space and $*_4$ is the Hodge dual.%
\footnote{Our conventions for the Hodge dual of a $p$-form in $D$-dimensions are
\begin{align*}
	*_D X_p  ~\equiv~ \frac{1}{p! (D-p)!}\, \epsilon_{m_1\ldots m_{D-p}, n_{D-p+1} \ldots n_D}\, X^{n_{D-p+1} \ldots n_D}\, e^{m_1}\wedge \ldots \wedge e^{m_{D-p}}\,,
\end{align*}
where $e^m$ denotes an orthonormal frame.}
This choice for $\beta$ and the base space metric, which we justify below, effectively reduces the remaining BPS equations to Laplace equations in four-dimensional flat space.
For the remaining layers, one can use the gauge symmetry of the theory to make convenient choices for the ansatz quantities \cite{Ceplak:2022wri}. 
However, we choose to work with gauge-invariant components of the  field strengths 
\begin{align}
	\label{eq:FieStrDef}
	F~=~ dA\,, \qquad G ~=~ dB + F\wedge A\,, \qquad 	-e^{2\phi} *_6 G ~=~ d\widetilde B\,,
\end{align}
where it is convenient to introduce an additional two form
\begin{align}
    \widetilde B ~& =~ -\frac{1}{Z_1}\, (du +\omega)\wedge (dv + \beta)+ a_1 \wedge (dv + \beta) + \gamma_2\,.
\end{align}
Using the form of the gauge field potentials and the definitions of the field strengths, one can show that the latter can be written as
\begin{subequations}
	\label{eq:Fields}
	\begin{align}
		F ~&  =~  (dv+ \beta) \wedge \omega_F + \tF\,,\\
		G ~&=~ d\left[ -\frac{1}{Z_2}\, (du +\omega)\wedge (dv + \beta)\right] + \widehat G_2\,,\\
		-e^{2\phi} *_6 G ~& = ~  d\left[ -\frac{1}{Z_1}\, (du +\omega)\wedge (dv + \beta)\right] + \widehat G_1\,.
	\end{align}
\end{subequations}
$\widehat G_{1,2}$ are two-forms on the  four-dimensional base space given by%
\footnote{In all expressions we have already assumed $\dot \beta =0$.}
	\begin{align}
		\widehat G_1 \equiv *_4 \cD Z_2 + (dv+ \beta)\wedge \Theta^1\,,\qquad 
		\widehat G_2  \equiv *_4 \cD Z_1  + (dv+ \beta)\wedge \Theta^2\,,
	\end{align}
where we used a differential operator
\begin{align}
    \cD  \equiv d_4 - \beta \wedge \pd_v\,.
\end{align}
The gauge-invariant forms $\Theta^{1,2}$, $\omega_F$, and $\tF$, that we introduced%
\footnote{$\Theta^2$ in this paper is denoted by $\widetilde \Theta^2$ in \cite{Ceplak:2022wri, Ceplak:2022pep}. The same difference in notation applies for $\cF$ and $a_2$.}  are given by 
\begin{align}
	\label{eq:U1Comp}
	\omega_F =-\dot \tA - \cD \left(\frac{Z_A}{Z_2}\right)\,,\qquad 
	\tF =- \cD \tA + \frac{Z_A}{Z_2}\, \, d_4 \beta\,,
\end{align}
and 
\begin{subequations}
	\begin{align}
		&\Theta^1 ~=~ \cD a_1 + \dot \gamma_2\,,
		&&*_4\cD Z_2 ~=~ \cD \gamma_2 - a_1 \wedge d_4\beta\,,\\
        %
        &\Theta^2 ~=~ \cD a_2  + \dot \gamma_1 + \tA \wedge \omega_F + \frac{Z_A}{Z_2}\, \tF \,,
		%
		&&*_4\cD Z_1  ~=~ \cD \gamma_1  - a_2  \wedge d_4\beta- \tF\wedge \tA\,. \label{eq:VectorInfluence}
	\end{align}
\end{subequations}
%

\paragraph{The first layer.}
%
The first non-trivial equations determine the pair $(Z_2, \Theta^1)$
	\begin{gather}
 	\label{eq:Lay1}
		*_4 \Theta^1   ~=~  \Theta^1 \,, \qquad
  \cD \Theta^1 ~=~   *_4 \cD \dot  Z_2 \,,
  \qquad\cD *_4 \cD Z_2 ~=~ -\Theta^1 \wedge d_4 \beta \,.
	\end{gather}
These equations are unchanged by the presence of the vector field: they only involve the degrees of freedom of the tensor multiplet.

\paragraph{The second layer.}
The next layer contains the information about the vector field 
	\begin{gather}
 	\label{eq:Lay2}
		*_4 \tF~=~ \tF\,,\qquad 
		2\, \cD Z_2 \wedge *_4\omega_F \,+\, Z_2\, \cD *_4 \omega_F~=~ - \tF \wedge \Theta^1\,.
	\end{gather}
While the self-duality condition can be solved independently, one cannot solve the second equation without first knowing the solutions to the first layer.
This is an explicit example of the upper-triangular structure of the BPS equations: The solutions to the first layer act as sources for the quantities appearing in the second layer.

\paragraph{The third layer.}
The next subset of equations determines $Z_1$ and $\Theta^2$:
	\begin{gather}
 \label{eq:Lay3}
		*_4 \Theta^2   =  \Theta^2\,,\qquad \cD\Theta^2 =   *_4 \cD \dot Z_1 - 2\, \omega_F \wedge \tF\,,\qquad 
		\cD *_4 \cD Z_1  =  \tF\wedge \tF-  \Theta^2 \wedge d_4 \beta  \,.
	\end{gather}
By comparing with \eqref{eq:Lay1}, we see the appearance of new quadratic source terms due to the presence of the vector fields. 
These terms break the symmetry between the pair $(Z_1, \Theta^2)$ and $(Z_2, \Theta^1)$ that is present in the system with only tensor multiplets \cite{Bena:2011dd,Giusto:2013rxa}.

\paragraph{The fourth layer.}
The last layer determines $\cF$ and $\omega$, which contain the information about the momentum along the $y$-direction and angular momentum in the $S^3$ respectively.
The two equations are 
\begin{align}
	\label{eq:Lay4Eq1}
	\cD \omega + *_4 \cD \omega + \cF \,d_4 \beta~=~   Z_1\, \Theta^1 + Z_2\,\Theta^2 \,,
\end{align}
and 
\begin{align}
	\label{eq:Lay4Eq2}
	*_4 \cD *_4   \left(\dot \omega - \frac12\, \cD \cF\right) = - \frac12 *_4 \Theta^1 \wedge\Theta^2 +  \ddot Z_1 \, Z_2 + \dot Z_1\, \dot Z_2 + Z_1 \, \ddot Z_2  + Z_2\,\left(\omega_F\right)_m\left(\omega_F\right)^m\,,
\end{align}
with the contraction in the terms involving $\omega_F$ being done with the base-space metric.

\section{Microstate geometries sourced by vector fields}
\label{sec:Gravity}

We now turn to the construction of the bulk dual of the state \eqref{eq:FullStateR}. 
We start by finding the perturbation around the maximally spinning supertube that is dual to the strands \eqref{eq:Rstate}.
Within six-dimensional supergravity this perturbation is encoded in a one-form gauge field, $A$.
By utilising the upper-triangular structure of the BPS equations,
we find the fully backreacted solutions, corresponding to smooth horizonless microstate geometries.
However, to obtain explicit solutions, one has to solve two partial differential equations whose closed-form solutions are not known. 
To remedy this, in the following section we provide a bootstrap-like method which in principle allows us to find all explicit solutions.


\subsection{Linearised solutions}

A simple solution to the BPS equations is the maximally spinning round supertube, whose dual CFT state is given in \eqref{eq:Supertubestate}.
One can think of this solution as the backreaction of the bound state of D1-branes and D5-branes which are distributed around a circle in $\mathbb{R}^4$ with radius $a$ .
It is convenient to introduce coordinates adapted to such a source
\begin{align}
    x_1 + i\,x_2 = \sqrt{r^2 +a^2}\,\sin\theta\,e^{i\phi}\,,\qquad x_3 + i\,x_4 = r\,\cos\theta\,e^{i\,\psi}\,,
\end{align}
with $\theta \in [0,\tfrac{\pi}{2}]$, $\phi, \psi \in [0,2\pi)$, and for now $a$ is taken to be an arbitrary constant.
In these coordinates the flat metric in $\mathbb{R}^4$ is given by 
\begin{align}
	\label{eq:BaseSpace}
	ds_4^2 = \Sigma\left(\frac{dr^2}{a^2 + r^2} + d\theta^2\right) + \left(a^2+ r^2\right)\, \sin^2\theta\, d\phi^2 + r^2 \, \cos^2\theta \, d\psi^2\,,  
\end{align}	 
where we introduced
\begin{align}
	\Sigma \equiv r^2 + a^2 \, \cos^2\theta\,.
\end{align}
Using these coordinates, one can express the ansatz quantities that define the maximally spinning round supertube as \cite{Lunin:2001fv, Lunin:2001jy}
\begin{subequations}
	\label{eq:Supertube}
	\begin{align}
		&Z_1 = \frac{Q_1}{\Sigma}\,,\qquad 
		Z_2 = \frac{Q_5}{\Sigma}\,,
		&&\gamma_{1,2} = - Q_{1,5} \frac{(r^2 + a^2)\cos^2\theta}{\Sigma}d\phi \wedge d\psi,\\
		&\beta=\frac{R_y\,a^2}{\sqrt{2}\,\Sigma}\,(\sin^2\theta\, d\phi - \cos^2\theta\,d\psi)\,,
		&&\omega=\frac{R_y\,a^2}{\sqrt{2}\,\Sigma}\,(\sin^2\theta\, d\phi + \cos^2\theta\,d\psi) \,,
	\end{align}
\end{subequations}
while all other ansatz quantities vanish.
The metric has a singularity at the location of the supertube, $\Sigma =0$,  unless one imposes
\begin{align}
\label{eq:RegularitySupertube}
	\sqrt{Q_1\, Q_5} = a \, R_y\,,
\end{align}
in which case the metric is smooth everywhere. 
This condition reduces the number of free parameters of the solution and can be seen as the gravitational equivalent of the condition \eqref{eq:StrandBudget} that constrains the total number of strands in the CFT state to be equal to $N$.

Now we would like to find the gravitational dual to the state \eqref{eq:FullStateR} with two types of strands.
Let us first consider the situation where $N_b \ll N_a$, as such a state corresponds to the supertube with an additional vector field perturbation.
The perturbation dual to the strands \eqref{eq:Rstate} is given by%
\footnote{The details of the construction of this perturbation is presented in Appendix~\ref{app:SolLay}.}
\begin{align}
	\label{eq:PertRkmn}
	A_{k,m,n} = \bt_{(k,m,n)}\, \Delta_{k,m,n}\, \left[\left( \frac{\sqrt{2}}{R_y}\, dv-d\phi - d\psi\right) \,\cos{\hat v_{k,m,n}}-  \frac{d\theta}{\sin\theta\cos\theta}\, \sin{\hat v_{k,m,n}} \right] \,.
\end{align}
In the above, $\bt_{(k,m,n)}\ll a$ is a parameter that measures the amplitude of the perturbation and $(k,m,n)$ are integers with 
\begin{align}
\label{eq:Ranges}
    k =2, 3,\ldots\,,\qquad m = 1,2,\ldots k-1\,,\qquad n =0,1,\ldots.
\end{align}
We also introduced
\begin{align}
	\Delta_{k,m,n} &\equiv
	\left(\frac{a}{\sqrt{r^2+a^2}}\right)^k
	\left(\frac{r}{\sqrt{r^2+a^2}}\right)^n 
	\cos^{m}\theta \, \sin^{k-m}\theta \,,
\end{align}
and 
\begin{align}
	\label{eq:vhatPhase}
	\hat{v}_{k,m,n} &\equiv (m+n) \frac{\sqrt{2}\,v}{R_y} + (k-m)\phi - m\psi \,.
\end{align}
If $(k,m,n)$ are taken to be outside the ranges in \eqref{eq:Ranges},  the perturbation becomes divergent at some point in the base space.

By comparing \eqref{eq:PertRkmn} with the supersymmetric ansatz \eqref{eq:Potentials}, we can read off the associated  quantities
\begin{subequations}
	\label{eq:PertAnsatz}
	\begin{align}
		Z_A^{(k,m,n)} &\equiv \frac{\sqrt{2}\,\bt_{(k,m,n)}\,Q_5}{R_y}\, \frac{\Delta_{k,m,n}}{\Sigma}\,\cos{\hat v_{k,m,n}}\,,\\
		\tA^{(k,m,n)} &\equiv  \bt_{(k,m,n)}\, \Delta_{k,m,n} \left[ \frac{d\theta}{\sin\theta\cos\theta}\, \sin{\hat v_{k,m,n}}  + \frac{(a^2 + r^2) \,d\phi + r^2\, d\psi}{\Sigma}\,\cos{\hat v_{k,m,n}} \right],
	\end{align}
\end{subequations}
up to $U(1)$ gauge transformations.
This ambiguity is irrelevant since in the BPS equations we  use the components of the gauge-field strength, $\omega_F$ and $\tF$.
These can be obtained using \eqref{eq:U1Comp}, and their explicit form is given in Appendix~\ref{app:SolLay} in \eqref{eq:OmFtF}.

One can check that the above perturbation solves the BPS equations to first order in $\bt_{(k,m,n)}/a\ll 1$. 
What is more, due to the linearity of the differential equations in \eqref{eq:Lay2}, any linear combination of \eqref{eq:PertRkmn} that sums over different $(k,m,n)$ is also a solution.
However, at first order in $\bt_{(k,m,n)}/a$ only the second layer equations are non-trivial, since in the subsequent layers $\omega_F$ and $\tF$ always appear quadratically and thus these equations are automatically satisfied at linear order in perturbation theory.
%

\subsection{Full, non-linear solutions}
\label{sec:Asy-AdS}

Going beyond first order in $\bt_{(k,m,n)}/a$  takes into account the backreaction of the perturbation and requires solving the BPS equations in full.
We focus on the case where only one mode (with fixed $(k,m,n)$) is excited, while leaving the extension to multi-mode superstrata for future work.
For simplicity, we  henceforth omit the subscripts in $\bt_{(k,m,n)}\equiv \bt$.

Let the perturbation be given by the single-mode solution \eqref{eq:PertAnsatz} and their field-strength components counterparts by \eqref{eq:OmFtF}.
If $\bt/a$ takes on a finite value, then one needs to solve the remaining BPS equations to all orders in this expansion parameter.
Already at second order, all fields can be excited, in principle. 
This includes the one-form $\beta$ and the base-space metric.
However, by looking at the layered structure of the BPS equations \eqref{eq:Lay1}-\eqref{eq:Lay4Eq2}, we note that the vector field components first appear in the second layer.
One can thus assume that by turning on a U(1) gauge field, one does not excite any fields whose components are determined by the layers preceding the second one. 
We thus assume that even when $\bt/a$ is finite, the base space metric, $\beta$, $Z_2$, and $\Theta^1$ do not change from \eqref{eq:Supertube} even with the presence of the vector field. 
This is somewhat justified by the fact that the BPS equations of the zeroth, first, and second layers are automatically satisfied for any value of $\bt$.
Then all that is left is to solve the final two layers. 

\subsubsection{Third layer}

In this layer the components of the vector field appear explicitly only through the quadratic terms $\tF \wedge \tF$ and $\omega_F\wedge \tF$.
Surprisingly, for single mode excitations, these terms are independent of the phase $\hat v_{k,m,n}$ and are thus $v$-independent.
The vector field also implicitly influences $\Theta^2$ and $Z_1$, as can be seen in \eqref{eq:VectorInfluence}.
But even the combinations appearing in those expressions are phase-independent.
Since there are no $v$-dependent source terms, we can assume $Z_1$ and $\Theta^2$ to be $v$-independent as well. 
In particular, we assume that $\dot Z_1 =\dot a_2= \dot \gamma_1 = 0$, which simplifies \eqref{eq:VectorInfluence} to 
	\begin{gather}
 \label{eq:Theta2vindAns}
		\Theta^2 ~=~ d_4a_2 + \tA \wedge \omega_F + \frac{Z_A}{Z_2}\, \tF  \,,\qquad 
		*_4 d_4 Z_1~=~ d_4 \gamma_1 - a_2 \wedge d_4\beta- \tF\wedge \tA\,.
	\end{gather}
These expressions automatically satisfy all BPS equations in the third layer apart from the $\Theta^2$ self-duality condition, which reduces to 
\begin{gather}
	\label{eq:a2eqNew}
	d_4 a_2 -  *d_4 a_2 = *\tA\wedge\omega_F - \tA\wedge\omega_F\,.
\end{gather}
The right-hand side of the above expression%
\footnote{For the explicit form see \eqref{eq:f1eqB}.}
is  $v$-independent, which is consistent with the assumption that $a_2$ is also independent of this coordinate, and contains no term proportional to $dr\wedge d\theta$, so one can make the following ansatz 
\begin{gather}
	\label{eq:Theta2ans}
	a_2 =\frac{\bt^2}{\sqrt2\,R_y}\biggr[ f_1(r,\theta)\left( d\phi + d\psi \right) + f_2 (r,\theta) \left( d\phi - d\psi\right)\biggr]\,.
\end{gather}
After inserting this  into \eqref{eq:a2eqNew} and some algebraic manipulations, one can express the derivatives of $f_1$ only in terms of $f_2$
\begin{subequations}
    \label{eq:f1eq}
    \begin{align}
        \pd_r f_1 &= \frac{2\,k\,r}{r^2 + a^2 \,\sin^2\theta}\, \Delta_{2k,2m,2n} + \frac{2\,r\,\sin\theta\,\cos\theta}{r^2 + a^2 \,\sin^2\theta}\,\pd_\theta f_2 + \left(1- \frac{2\,r^2\,\cos^2\theta}{r^2 + a^2\,\sin^2\theta}\right)\,\pd_r\,f_2\,,\label{eq:f1req}\\
        \pd_\theta f_1 &= \left(1- \frac{2\,r^2\,\cos^2\theta}{r^2 + a^2\,\sin^2\theta}\right)\,\pd_\theta f_2 - 2\,\frac{\left(a^2 +r^2\right)\,r\,\sin\theta\,\cos\theta}{r^2 + a^2\,\sin^2\theta}\,\pd_r f_2\nonumber\\*
        &\quad -\frac{2 \tan\theta}{r^2 + a^2\,\sin^2\theta}\left(k\,r^2\,\cot^2\theta + n\,\frac{r^2 + a^2\,\sin^2\theta}{\sin^2\theta}\right)\,\Delta_{2k,2m,2n}\,\label{eq:f1theq},
    \end{align}
\end{subequations}
while $f_2$ is determined by a Laplace equation on the four-dimensional base
\begin{equation}
	\label{eq:Lap1}
	\cLh f_2 = -\frac{2\,a^2}{\Sigma(a^2 + r^2)\, \cos^2\theta}\,\Big[ (k+n)^2\, \Delta_{2k,2m+2,2n}+ n^2\, \Delta_{2k,2m,2n-2}  -k(m+n)\, \Delta_{2k-2, 2m, 2n}\Big],
\end{equation}
with $\cLh$ being given by
\begin{align}
	\label{eq:LapDef}
	\cLh f(r,\theta) \equiv \frac{1}{r\, \Sigma}\,\pd_r \left(r\left(a^2 + r^2\right)\pd_r f\right)+ \frac{1}{\Sigma\, \cos\theta\, \sin\theta}\,\pd_\theta\left(\cos\theta\, \sin\theta\pd_\theta f\right)\,.
\end{align}
Impressively, the differential equation \eqref{eq:Lap1} admits a closed form solution \cite{Bena:2015bea,Bena:2017xbt}
\begin{align}
	\label{eq:f2sol}
	f_2^{(k,m,n)}(r,\theta) &= -2 \,a^2\Big[(k+n)^2\, F_{2k,2m+2,2n} + n^2\, F_{2k,2m,2n-2}   - k\,(m+n)\, F_{2k-2, 2m, 2n} \Big]\,,
\end{align}
where 
\begin{equation} \label{eq:Ffun}
	F_{2k,2m,2n}=-\!\sum^{j_1+j_2+j_3\le k+n-1}_{j_1,j_2,j_3=0}\!\!{j_1+j_2+j_3 \choose j_1,j_2,j_3}\frac{{k+n-j_1-j_2-j_3-1 \choose k-m-j_1,m-j_2-1,n-j_3}^2}{{k+n-1 \choose k-m,m-1,n}^2}
	\frac{\Delta_{2(k-j_1-j_2-1),2(m-j_2-1),2(n-j_3)}}{4(k+n)^2(a^2 + r^2)}\,,
\end{equation} 
and
\begin{equation} 
	{j_1+j_2+j_3 \choose j_1,j_2,j_3}\equiv \frac{(j_1+j_2+j_3)!}{j_1!\, j_2!\, j_3!}\,.
\end{equation} 

Once $f_2$ is calculated, one can in principle find a closed-form expression for $f_1$ using hypergeometric functions. 
However, in practice it is much easier to integrate the first-order differential equations \eqref{eq:f1eq}  case-by-case. 
Once $f_1$ and $f_2$ are known, one has all the ingredients to calculate  $\Theta^2$ and thus explicitly solve the third layer BPS equations \eqref{eq:Lay3}.

Let us conclude this part by noting that one can solve all BPS equations in this layer without explicitly determining $Z_1$. 
Namely, when $Z_1$ is $v$-independent, it appears in the BPS equations only through $*_4 d_4 Z_1$, which can be expressed by other known quantities using \eqref{eq:VectorInfluence}.
One can solve even the fourth layer equations without ever calculating $Z_1$: When $\dot Z_1 =0$, this ansatz quantity appears only in \eqref{eq:Lay4Eq1} where it multiplies $\Theta^1$, which we assume to be vanishing.
Solving for $Z_1$ is nonetheless important as it crucially effects the asymptotic behaviour and the regularity of the geometry. 
When $\dot Z_1 =0$, the rightmost equation of \eqref{eq:Lay3} can be rearranged to%
\footnote{When acting on a scalar function, the Laplace operator is defined as $\cLh F \equiv - *_4 d_4*_4d_4 F$.}
\begin{align}
	\label{eq:Z1Lap}
	\cLh Z_1 &=  *_4\left(\Theta^2 \wedge d_4 \beta - \tF \wedge\tF\right) = \frac{\,a^2\, \bt^2}{(r^2+ a^2
	 		\sin^2\theta)}\left[\cLh\left(\frac{f_2}{\Sigma}\right) - \frac{1}{\Sigma}\,\cLh f_2\right]\nonumber\\*
 %
		%
		&\quad  +\frac{4 \,a^4\,\bt^2\,k}{\Sigma^3\,(r^2+ a^2
			\sin^2\theta)}\,\Delta_{2k,2m,2n} -  \frac{2\,\bt^2}{\Sigma^2}\, \frac{\Delta_{2k, 2m, 2n-2}}{ \cos^2\theta\,\sin^2\theta}\,\left(\frac{n\,a^2- k\,r^2}{a^2+r^2}\right)^2\,.
\end{align}
We are currently lacking an analytic solution to this equation for all values of $(k,m,n)$. 
This is mainly due to two complicating factors.
Firstly, we notice that in order to find $Z_1$ one already needs to know $f_2$. 
But from \eqref{eq:f2sol}, we see that this function becomes more complicated as the $(k,m,n)$ are increased. 
The differential equation \eqref{eq:Z1Lap} can thus have an arbitrary number of terms, unlike \eqref{eq:Lap1} which always has only three source terms on the right-hand side.
Secondly, while the source terms in \eqref{eq:Lap1} are singular at $\Sigma= 0$, this diverging behaviour is cancelled out by an analogous factor in the Laplace operator \eqref{eq:LapDef}.
In \eqref{eq:Z1Lap}, the most singular terms scale as $\Sigma^{-3}$, which cannot be fully cancelled out by $\cLh$. 
This stronger singular behaviour comes from the $Z_1$ function itself: Unlike $f_2$ that is regular everywhere, $Z_1\sim \Sigma^{-1}$, and thus diverges at the supertube location.
The singular source terms in the above differential equation are manifestations of the brane sources being located at the supertube locus $\Sigma =0$.

\subsubsection{Fourth layer}

Since $\Theta^1 =\dot Z_1 = \dot Z_2 =0$, we see that there is only one non-vanishing source term in the fourth layer BPS equations \eqref{eq:Lay4Eq1} and \eqref{eq:Lay4Eq2} 
\begin{align}
	Z_2 \, \omega_F^2 = \frac{2 \,\bt^2\, Q_5}{R_y^2}\, \frac{1}{{\Sigma\,(a^2 + r^2)\, \cos^2\theta}}\Big[n^2\, \Delta_{2k,2m,2n-2} + (k+n)^2\, \Delta_{2k,2m+2,2n}\Big]\,,
\end{align}
which is again $v$-independent.
We can then assume that $\cF$ and $\omega$ are also $v$-independent, $\dot \cF = \dot \omega=0$,%
\footnote{The more complicated alternative is that both $\omega$ and $\cF$ are $v$-dependent, but their dependence cancels out in such a way that the right-hand sides of both  \eqref{eq:Lay4Eq1} and \eqref{eq:Lay4Eq2} are $v$-independent.}
which decouples the two differential equations. 
Equation \eqref{eq:Lay4Eq2} reduces to 
\begin{align}
	\cLh\,\cF =\frac{4 \,\bt^2\, Q_5}{R_y^2}\, \frac{1}{{\Sigma\,(a^2 + r^2)\, \cos^2\theta}}\Big[(k+n)^2\,\Delta_{2k,2m+2,2n} + n^2\, \Delta_{2k,2m,2n-2}\Big]\,,
\end{align}
which is solved by
\begin{align}
	\label{eq:cFsol}
	\cF^{(k,m,n)} =  \frac{4\, \bt^2\, Q_5}{R_y^2}\Big[(k+n)^2\,F_{2k,2m+2,2n} + n^2\, F_{2k,2m,2n-2}\Big]\,.
\end{align}
The terms in the square bracket are exactly the first two terms appearing in the square bracket in \eqref{eq:f2sol}, but we do not have a good explanation for why this is the case.

Once $\cF$ is determined, one only needs to solve
\begin{align}
	\label{eq:OmegaeqkmnScheme}
	d_4 \omega + *_4 d_4\omega &= Z_2 \, \Theta^2 - \cF \, d_4\beta\,.
\end{align}
The right-hand side, which we give explicitly in \eqref{eq:Omegaeqkmn}, is $v$-independent and does not contain a $dr\wedge d\theta$ term and so we can make an ansatz for the $\bt^2$ correction to $\omega$ as
\begin{align}
	\label{eq:OmegaBAns}
	\omega_{\bt} = \frac{\sqrt2\,Q_5\,\bt^2}{R_y} \Big[\mu(r,\theta)\, \left( d\phi + d\psi \right) + \nu (r,\theta) \left( d\psi-d\phi \right)\Big]\,.
\end{align}
After inserting this ansatz into \eqref{eq:OmegaeqkmnScheme} and some algebraic manipulation,  one can express the partial derivatives of $\nu$ in terms of $\mu$, $f_2$, and $\cF$, see \eqref{eq:nuEq}.
Eliminating $\nu$ in this manner leaves us with a single Laplace equation for $\mu$
\begin{align}
	\label{eq:muEq}
	&\cLh\left(\mu + \frac{r^2+ a^2\,\sin^2\theta}{2\, \Sigma}\,f\right) =\frac{a^2}{4\,(r^2+ a^2\,\sin^2\theta)}\,\left[\cLh\left(\frac{f_2}{\Sigma}\right) - \frac{1}{\Sigma}\,\cLh f_2\right]\nonumber\\* 
    &+ \frac{n^2\,a^4}{2\,\Sigma\,\left(a^2+ r^2\right)^2}\,\Delta_{2k,2m,2n-2}- \frac{k\,(k+2n)\,a^2}{2\,\Sigma\,\left(a^2+ r^2\right)}\,\Delta_{2k,2m,2n}- \frac{k\,a^4}{\Sigma^2\,\left(r^2+a^2\,\sin^2\theta\right)}\,\Delta_{2k,2m,2n}\nonumber\\* 
    &+\frac{n^2\,a^2 -  (2\,k\,n+n^2)\,r^2}{2\,\Sigma\,\left(a^2+r^2\right)\,\cos^2\theta}\,\Delta_{2k,2m,2n-2} + \frac{(k+n)\left((k-n)\,r^2- 2\,n\,a^2\right)}{2\,\Sigma\,\left(a^2+r^2\right)\,\cos^2\theta}\,\Delta_{2k,2m+2,2n}\,,
\end{align}
where we have defined 
\begin{align}\label{eq:cFexp}
	\cF \equiv  \frac{4\, \bt^2\, Q_5}{R_y^2}\,f(r,\theta)\,.
\end{align}
Once $\mu$ is determined, one can find $\nu$ by  integrating the differential equations \eqref{eq:nuEq}.

We are unable to solve \eqref{eq:muEq} with a closed form expression for all possible values of $(k,m,n)$. The reasons are the same as with \eqref{eq:Z1Lap} -- the explicit appearance of $f_2$ and the stronger divergence of $\mu$ near $\Sigma = 0$.
However, we are able to solve this equation and \eqref{eq:Z1Lap} for some specific values of $(k,m,n)$: This shows that these differential equations have solutions, only that finding them is non-trivial. 
In light of this, we present in the next section an algorithm with which one can in principle calculate these solutions for any $(k,m,n)$. 
While we are unable to provide a solution that would be similar in quality as \eqref{eq:cFsol}, we automate the procedure of finding individual solutions.

\section{Explicit examples}
\label{sec:Bootstrapy}
\subsection{Bootstrapping the result}

Here we outline a method to `bootstrap' the solutions to the BPS equations. We implement this method to compute both $\mu^{(k,m,n)}$ and  $Z^{(k,m,n)}_1$ for arbitrary fixed $(k,m,n)$.

\subsubsection{Bootstrapping $\mu^{(k,m,n)}$}\label{mu bootstrap}

Here we bootstrap for $\mu^{(k,m,n)}$, which solves the partial differential equation 
\begin{align}
	\label{eq:muEq2}
	&\cLh\left(\mu^{(k,m,n)} + \frac{r^2+ a^2\,\sin^2\theta}{2\, \Sigma}\,f^{(k,m,n)}\right) =\frac{a^2}{4\,(r^2+ a^2\,\sin^2\theta)}\,\left[\cLh\left(\frac{f_2^{(k,m,n)}}{\Sigma}\right) - \frac{1}{\Sigma}\,\cLh f_2^{(k,m,n)}\right]\nonumber\\* 
    &+ \frac{n^2\,a^4}{2\,\Sigma\,\left(a^2+ r^2\right)^2}\,\Delta_{2k,2m,2n-2}- \frac{k\,(k+2n)\,a^2}{2\,\Sigma\,\left(a^2+ r^2\right)}\,\Delta_{2k,2m,2n}- \frac{k\,a^4}{\Sigma^2\,\left(r^2+a^2\,\sin^2\theta\right)}\,\Delta_{2k,2m,2n}\nonumber\\* 
    &+\frac{n^2\,a^2 -  (2\,k\,n+n^2)\,r^2}{2\,\Sigma\,\left(a^2+r^2\right)\,\cos^2\theta}\,\Delta_{2k,2m,2n-2} + \frac{(k+n)\left((k-n)\,r^2- 2\,n\,a^2\right)}{2\,\Sigma\,\left(a^2+r^2\right)\,\cos^2\theta}\,\Delta_{2k,2m+2,2n}\,.
\end{align}
Recall that the expression for $f_2(r,\theta)$ is given in (\ref{eq:f2sol}) while $f(r,\theta)$ is given in the square bracket in \eqref{eq:cFexp}.
Though the source term, at first glance, appears to be quite complicated, for fixed $(k,m,n)$ it reduces to a polynomial function in the numerator composed of trigonometric functions of $\theta$ and powers of $r^2$ with the denominator also containing factors of $\Sigma$ and $(r^2 + a^2\,\sin^2\theta)$.
We use this to make an informed ansatz for the solution.

Based on the form of known six-dimensional superstrata \cite{Bena:2015bea, Bena:2016agb, Bena:2016ypk, Bena:2017xbt, Ceplak:2018pws, Heidmann:2019zws, Heidmann:2019xrd, Shigemori:2020yuo,  Ceplak:2022pep}, the various components of the solutions for $\mu$ can be written as combinations of
\bea 
r^{2},\quad \cos^2\theta, \quad {1\over (r^2+a^2)}\,,
\eea
raised to some non-negative powers.
Using these functions ensures that the resulting solutions are smooth at potentially singular locations. Namely, these factors are regular at the origin ($r=0$ and $\theta=0$) and the supertube location ($r=0$ and $\theta = \pi/2$).
Nonetheless, the most singular term in \eqref{eq:muEq2} comes from the first line and diverges as $\Sigma^{-3}$ near the supertube location, which suggests that $\mu^{(k,m,n)}$ should scale as $\Sigma^{-1}$. 
These considerations lead us to the following ansatz
\bea\label{ansatz} 
\mu^{(k,m,n)} = C^0_{k,m,n} + {C^H_{k,m,n}\over\Sigma} +{1\over\Sigma}\sum_{k_1=0}^n\sum_{k_2=0}^{k}\sum_{k_2=0}^{k+n}C_{k_1,k_2,k_3}{(r^{2})^{k_1}(\cos^{2}\theta)^{k_2}\over(r^2+a^2)^{k_3}}
\eea 
where we have added an arbitrary constant $C^0_{k,m,n}$ and homogeneous term ${C^H_{k,m,n}}/{\Sigma}$. This ansatz contains three sums, which is tractable for low mode numbers, 
but is computationally too demanding for higher $(k,m,n)$.
There are also redundant coefficients, which end up vanishing. 
To remedy this, we reduce the number of sums by rewriting the ansatz with a common denominator, shifting the indices of the components in the numerator
\bea\label{ansatz prime} 
\mu^{(k,m,n)} = C^0_{k,m,n} + {C^H_{k,m,n}\over\Sigma}+ {1\over\Sigma}\sum_{k_1=0}^{k+n}\sum_{k_2=0}^{k}C_{k_1,k_2}{a^{2k+kn}(r^{2}/a^2)^{k_1}(\cos^{2}\theta)^{k_2}\over(r^2+a^2)^{k+n}}
\eea 
which is just an expansion of the initial ansatz in a different basis.
To solve the differential equation in practice, we introduce 
dimensionless quantities by setting $a=1$ and defining
\begin{align}
r&=\sqrt{z-1},\quad\cos\theta= \sqrt y\,,
\end{align}
in which case \eqref{ansatz prime} reads
\begin{equation}\label{ansatz prime 2} 
\mu^{(k,m,n)} = C^0_{k,m,n} + {C^H_{k,m,n}\over\Sigma}+ {1\over\Sigma}{1\over  z^{k+n}}\bigg(\sum_{k_1=0}^{k+n}\sum_{k_2=0}^{k}C_{k_1,k_2}(z-1)^{k_1}y^{k_2} -(z-1)^{k+n}C_{k+n,0} \bigg),
\end{equation} 
where we have isolated an explicit homogeneous term in the sum, and $\Sigma = y+z-1$.
We then rewrite the source term and the Laplace operator in \eqref{eq:muEq2} using these coordinates. 
This transforms the partial differential equation into a set of algebraic equations for the coefficients $C_{k_1,k_2}$.
%
This algorithm can determine the solution to $\mu^{(k,m,n)}$ up to the constant and homogeneous terms, since these vanish under the action of the Laplace operator. 
They are determined by the regularity condition of the metric, which we discuss in Section~\ref{rc}.

\subsubsection{Bootstrapping $Z^{(k,m,n)}_1$} 
In order to completely specify the metric we must find the explicit form of $Z_1$. We again use the `bootstrap' method to solve 
\begin{align}
\label{eq:Z1eq2}
\cLh Z_1^{(k,m,n)} &= \frac{4\,a^2\, \bt^2}{\Sigma^3\,(r^2+ a^2
			\sin^2\theta)}\left(\,a^2 \,\sin\theta\,\cos\theta \,\pd_{\theta} f_2^{(k,m,n)} -  r(a^2 + r^2) \pd_r f_2^{(k,m,n)}\right)\nonumber\\
		&\quad +\frac{4 \,a^4\,\bt^2\,k}{\Sigma^3\,(r^2+ a^2
			\sin^2\theta)}\,\Delta_{2k,2m,2n} -  \frac{2\,\bt^2}{\Sigma^2}\, \frac{\Delta_{2k, 2m, 2n-2}}{ \cos^2\theta\,\sin^2\theta}\,\left(\frac{n\,a^2- k\,r^2}{a^2+r^2}\right)^2\,.
\end{align}
Following the above discussion for $\mu^{(k,m,n)}$, we make a 
similar dimensionless ansatz
\bea 
\label{eq:Z1BAns}
Z^{(k,m,n)}_1 = D^0_{k,m,n} + {D^H_{k,m,n}\over\Sigma} + {1\over\Sigma}\sum_{k_1=0}^{k+n-1}\sum_{k_2=0}^{k}D^{k,m,n}_{k_1,k_2}(z-1)^{k_1}y^{k_2},
\eea
where we again pick the summation range appropriately. 
Inserting this ansatz into \eqref{eq:Z1eq2} yields another set of algebraic equations that can be solved. 
As before, the homogeneous and constant terms remain undetermined using this procedure. We fix them by demanding that the function  $Z_1^{(k,m,n)}$ has an appropriate asymptotic fall-off.

\subsubsection{Finding the remaining ansatz quantities: $\nu^{(k,m,n)}$ and $f_1^{(k,m,n)}$}

While $\mu^{(k,m,n)}$ and $f_2^{(k,m,n)}$ are determined using second order partial differential equations, their counterparts, $\nu^{(k,m,n)}$ and $f_1^{(k,m,n)}$ respectively, are determined by first order differential equations. 
Once $\mu^{(k,m,n)}$ and $f_2^{(k,m,n)}$ are found for fixed $(k,m,n)$ it is straightforward to integrate the necessary first order differential equations.
In particular, to obtain  $\nu^{(k,m,n)}$ we take the $\mu^{(k,m,n)}$ found using the method presented in  Section~\ref{mu bootstrap} for a specified $(k,m,n)$ and insert these solutions into (\ref{eq:nuEq}), which can be integrated.
This determines $\nu^{(k,m,n)}$ up to $C^0_{k,m,n}$, $C^H_{k,m,n}$, and $\bar C^0_{k,m,n}$ an additional constant one picks up during the integration of the differential equations determining $\nu$.  These constants are then determined by regularity conditions in the metric. 
In order to determine the form $a^{(k,m,n)}_2$ we need to compute $f_1^{(k,m,n)}$. To do this we follow a similar procedure as for $\nu^{(k,m,n)}$. We take the expression computed for $f^{(k,m,n)}_2$, (\ref{eq:f2sol}), for a given $(k,m,n)$, and insert it into (\ref{eq:f1eq}), which can be easily integrated. This procedure  both $f_1^{(k,m,n)}$ and $f_2^{(k,m,n)}$ up to a constant factor which is irrelevant as it is pure gauge.

\subsubsection{Regularity conditions}\label{rc}

The explicit solutions for $\mu^{(k,m,n)}$, $\nu^{(k,m,n)}$  and $Z_1^{(k,m,n)}$ are obtained by solving differential equations and so there are constant and homogeneous terms which remain undetermined.
These  are  fixed by requiring that the solutions have the right asymptotic behaviour and are regular (without singularities or closed timelike curves) in the interior. 
From the supersymmetric ansatz \eqref{eq:MetAns1}, one can see that the near-boundary behaviour of the combination $Z_1\, Z_2$ determines the asymptotic radius of AdS$_3\times S^3$.
We want the asymptotic radius to remain unchanged -- equal to $R_{\rm AdS}^2 = \sqrt{Q_1\,Q_5}$. Because $Z_2$ is the same as in the supertube geometry, this imposes the condition on the large-$r$ fall off of $Z_1$
\begin{align}
    \label{eq:Z1FallOff}
    Z_1^{(k,m,n)} \xrightarrow[r\to \infty]{} \frac{Q_1}{r^2}+ \cO\left(r^{-3}\right)\,,
\end{align}
which uniquely fixes the coefficient $D^0_{k,m,n}$  and $D^H_{k,m,n}$.

Next, we look at the behaviour of $\omega$.
The components of this one-form contain information about the angular momentum of the geometry  \cite{Myers:1986un, Giusto:2013bda}
\begin{align}
\label{eq:AngularMomentum}
    \beta_{\phi} + \beta_{\psi}+ \omega_{\phi} + \omega_{\psi}\xrightarrow[r\to \infty]{}\sqrt{2}\,\frac{J- \Jb\,\cos2\theta}{r^2}\,.
\end{align}
While the extraction of  $J$ and $\Jb$ can be done on a case-by-case basis, we note that $\omega$ cannot have any terms which fall-off slower at the asymptotic boundary than $r^{-2}$. This directly imposes a condition on the large-$r$ behaviour of $\mu^{(k,m,n)}$
\begin{align}
    \label{eq:OmegaCond1}
    \mu^{(k,m,n)}\xrightarrow[r\to \infty]{} \cO\left(r^{-2}\right)\,.
\end{align}
Imposing this condition uniquely fixes the value of the constant term $C^0_{k,m,n}$.
It also ensures that the spacetime asymptotes to AdS$_3\times S^3$ and not a deformation of it.

In order for the solutions to be regular in the interior, the components of $\omega$ should remain finite (in a non-degenerate coordinate system).
In our four-dimensional base, there are two potentially singular points, the supertube locus $\Sigma =0$, and the origin of spacetime: $r=0$, $\theta =0$.
Following \cite{Bena:2017xbt}, we first impose that the $(d\phi + d\psi)$ and $(d\phi-d\psi)$ components of $\omega$ vanish at the origin.%
\footnote{By using the combination of angles $\varphi_{\pm} =\phi \pm \psi$ we can rewrite the flat $\mathbb{R}^4$ metric in Gibbons-Hawking form. The condition that the $d\varphi_{\pm}$ components of $\omega$ vanish at the origin can then be seen as a regularity condition for the sizes of the $\varphi_{\pm}$ circles to vanish at the origin of Gibbons-Hawking space.}
This in turn demands that near the origin, the $\mu^{(k,m,n)}$ and $\nu^{(k,m,n)}$ terms vanish:
\begin{align}
    \label{eq:munucond}
    \mu^{(k,m,n)} \xrightarrow[~r\to 0~]{~\theta \to 0~} 0\,, \qquad \nu^{(k,m,n)} \xrightarrow[~r\to 0~]{~\theta \to 0~} 0\,.
\end{align}
Imposing these two conditions determines the coefficients $C^H_{k,m,n}$ and  $\bar C^0_{k,m,n}$.

The final condition relates $Q_1$, $Q_5$, $a$, $\bt$, and $R_y$.
Recall that for the supertube solution, regularity near the $\Sigma =0$ locus imposed \eqref{eq:RegularitySupertube} which related the four parameters of the solution. 
To establish the regularity condition for the new solutions, one again needs to look at the solutions near the supertube locus: 
In particular, their $(d\phi + d\psi)^2$ and $(d\phi - d\psi)^2$ components of the metric.
It is convenient to define
\begin{align}
     r = a\,\lambda\, \cos\chi\,, \qquad \theta = \frac{\pi}{2}- \lambda\,\sin\chi\,,\label{eq:STLim}
\end{align}
as in these coordinates  $\Sigma = a^2\,\lambda^2+ \cO(\lambda^4)$,
and thus sending $\lambda \to 0$ corresponds to $\Sigma \to 0$.
To extract the regularity condition, we find it convenient  to write the backreacted ansatz quantities in an expansion in $\lambda$
\begin{subequations}
\label{eq:RegAnsatz}
    \begin{gather}
        Z_1 = \frac{Q_1}{\Sigma}+ \bt^2\left[\frac{\delta Z_1}{a^2\,\lambda^2}+ \coo{\lambda^0}\right]\,, \\
        \mu = \frac{\delta \mu}{a^2\,\lambda^2}+ \coo{\lambda^0}\,,\qquad \nu = \frac{\delta \nu}{a^2\,\lambda^2}+ \coo{\lambda^0}\,,\qquad f = \delta f + \coo{\lambda^2}\,,
    \end{gather}
\end{subequations}
which is straightforward once these quantities have been calculated using the above procedure.
One then inserts these expansions, together with the values for the unperturbed supertube \eqref{eq:Supertube}, into the metric \eqref{eq:MetAns1} and extracts the relevant metric coefficients.
The metric is regular at the supertube locus, if there are no terms which diverge as $\lambda\to 0$.
This happens when the following two constraints are met
    \begin{align}
    \label{eq:GenRegularity}
        Q_1\,Q_5 = a^2\,R_y^2 + \bt^2\,Q_5\left(2\,a^2 \,\delta f - \delta Z_1 + 4\,\delta\mu\right)\,,\qquad   \delta \mu = -\delta \nu\,,
    \end{align}
the latter of which is just the condition that the components of the one-form $\omega$ have the same divergent behaviour near the supertube locus.
The constraint \eqref{eq:GenRegularity} replaces the regularity condition for the supertube -- indeed, by setting $\bt = 0$ we recover \eqref{eq:RegularitySupertube}.
Analysing \eqref{eq:GenRegularity} for a few explicit examples, one obtains
\begin{align}
    Q_1 \,Q_5 = a^2\,R_y^2 + \frac{x_{k,m,n}}{2}\,\bt^2\,Q_5\,,
    \label{eq:ExplicitRegularity0}
\end{align}
where
\begin{align}
    x_{k,m,n} \equiv k\,\frac{(k-m-1)!\,(m-1)!\,n!}{(k+n-1)!}\,.
\end{align}
It is more convenient to rescale $\bt$ as
\begin{align}
    b^2 \equiv \bt^2 \frac{Q_5}{R_y^2}\,,
\end{align}
in which case the regularity condition can be recast as
\begin{align}
    Q_1 \,Q_5 = a^2\,R_y^2 + \frac{x_{k,m,n}}{2}\,b^2\,R_y^2\,,
    \label{eq:ExplicitRegularity}
\end{align}
which is more familiar from previously known examples.

\subsubsection{Asymptotic charges}

A non-trivial check of the claim that the supergravity solutions constructed above are dual to the CFT states \eqref{eq:FullStateR}, is to compare the asymptotic charges that can be read off from the geometry with the corresponding charges of the CFT state. 
The starting point is to compare \eqref{eq:ExplicitRegularity} with the corresponding strand budget condition \eqref{eq:ActualConstraint} and determine that 
\begin{align}
    \frac{N_a}{N} \equiv \frac{a^2\,R_y^2}{Q_1\,Q_5}\,,\qquad \frac{N_b}{N} = \frac{x_{k,m,n}}{2\,k}\,\frac{b^2\,R_y^2}{Q_1\,Q_5}\,.
    \label{eq:Translation}
\end{align}

Next, we need to read off the asymptotic charges from the fully backreacted geometries. 
Here we take a shortcut: Instead of analysing the conserved charges in the asymptotically AdS$_3\times S^3$ solutions, we assume that these geometries can be coupled to flat space and in doing so their asymptotic behaviour does not change. 
In particular, we assume that the only non-trivial change is that the harmonic functions $Z_I$ obtain an additional constant term that ensures that the geometry is asymptotically flat, that is $\mathbb{R}^{4,1}\times S^1_y$.
Then we can use the standard expressions for determining conserved charges in asymptotically flat solutions, see for example \cite{Warner:2019jll}.
We justify this simplification in two ways. 
From a physical point of view, in the geometries we constructed the non-trivial deformation away from pure AdS$_3\times S^3$ is localised in the interior of the solution, with all associated ansatz quantities having a well-defined asymptotic fall-off, as in \eqref{eq:Z1FallOff} or \eqref{eq:OmegaCond1}.
Therefore, coupling such solutions to flat space should not excite any $v$-independent terms that have a slower fall-off at the boundary of AdS, similar to what was observed in \cite{Bena:2017xbt, Ceplak:2018pws}.
Secondly, as we show below, we find that the conserved charges obtained in this way agree perfectly with the CFT predictions, retrospectively justifying our choice of method.

Let us begin by analysing the momentum, which can be read off from the large $r$ fall-off of the  $\cF$ function as
\begin{align}
    \cF = -\frac{2\,Q_P}{r^2}+ \coo{r^{-3}}\,.
\end{align}
Since $\cF$ is known in closed form for generic $(k,m,n)$, \eqref{eq:cFsol}, we can read-off
\begin{align}
\label{eq:MomRes}
    Q_P = \frac{b^2}{2}\,\frac{m+n}{k}\,x_{k,m,n}\,.
\end{align}
Next, we analyse the angular momentum, which is encoded in the one-forms $\beta$ and $\omega$, as given in \eqref{eq:AngularMomentum}.
Here we lack an analytic result, but by examining several examples one can see that 
\begin{align}
\label{eq:AngMomRes}
    J = \frac{a^2\,R_y}{2} + \frac{b^2\,R_y}{2}\,\frac{m}{k}\,x_{k,m,n}\,,\qquad \Jb = \frac{a^2\,R_y}{2}\,.
\end{align}
A judicious check is that $\Jb$ is unchanged from the value of the round supertube solution, since we are only acting on the left sector of the CFT and thus any conserved quantities associated with the right sector should remain the same.

If the geometries that we analysed above are considered in the D1-D5 frame%
\footnote{See \cite{Ceplak:2022pep} for how to uplift the six-dimensional solutions to solutions of ten-dimensional supergravity in different duality frames.}
then one can compare the supergravity charges with the eigenvalues of the CFT operators discussed in Section~\ref{sec:CFT}.
In particular, the dictionary relating the supergravity charges $Q_P$, $J$, and $\Jb$ to the quantised charges $n_P$, $j$, and $\jb$ is \cite{Bena:2017xbt}
\begin{align}
 n_P =  \frac{N\, R_y^2}{Q_1\, Q_5}\, Q_P\,,\qquad j =  \frac{N\, R_y}{Q_1\,Q_5}\, J\,,\qquad   \jb =  \frac{N\, R_y}{Q_1\,Q_5}\, \Jb\,.
\end{align}
Together with the identification \eqref{eq:Translation}, one obtains
\begin{align}
    n_P = N_b\,(m+n)\,, \qquad j = \frac{N_a}{2} + N_b\,m\,, \qquad \jb = \frac{N_a}{2}\,,
\end{align}
which indeed  matches the CFT results in \eqref{eq:FullRCharges} and \eqref{eq:MomChargeCFT}. 

\subsection{An analytic family of solutions}

In principle, with the procedure presented above, we are able to find explicit solutions with arbitrary numbers $(k,m,n)$.
As an example, we present the solution with $(k=4,m=2,n=3)$ in full detail in Appendix~\ref{app:example}. 
However, the simplest family of solutions, labelled by $(k=2,m=1,n)$, can be written in a relatively concise form and it may be worthwhile to examine these geometries in more detail to get a better feel about their properties.
The ansatz quantities for the $(2,1,n)$ geometries are given by
\begin{subequations}
    \label{eq:21nSol}
    \begin{align}
        Z_1^{(2,1,n)} &= \frac{Q_1}{\Sigma} - \frac{\tilde b^2}{2\,\Sigma}\,\frac{a^4}{\left(a^2+r^2\right)^2}\,\Delta_{0,0,2n}\,,\\
        f^{(2,1,n)} &= \frac{r^2 +a^2\,\sin^2\theta}{4\,\left(a^2+r^2\right)^2}\,\Delta_{0,0,2n}-\frac{1}{2\,(n+1)\,a^2}\,\left(1-\Delta_{0,0,2n+2}\right)\,,\\
		 a_2^{(2,1,n)}&= \frac{a^2\,\tilde b^2}{\sqrt2\,R_y\left(a^2 +r^2\right)}\,\Delta_{0,0,2n}\,\left(\frac{r^2\,\cos^2\theta}{a^2 +r^2}\,d\psi-\sin^2\theta\, d\phi\right)\,,\\
		\mu^{(2,1,n)} &=\frac{1}{4\,(n+1)\Sigma}\Bigg[1-\cos(2\theta)\left(1-\frac{a^2(n+1)+r^2}{a^2 +r^2}\,\Delta_{0,0,2n}\right)-\frac{(n+1)\,a^2}{a^2 + r^2}\,\Delta_{2,2,2n}\Bigg] \,,\\
		\nu^{(2,1,n)} &=\frac{1}{4\,(n+1)\Sigma}\Bigg[\cos(2\theta)-1+\left(1+ \frac{n\,a^2}{a^2 +r^2}\right)\Delta_{0,0,2n}-\frac{(n+1)\,a^2}{a^2+r^2}\, \Delta_{2,2,2n}\Bigg]\,,\\
	\gamma_1^{(2,1,n)} &= - Q_1\,\frac{(a^2+r^2)\,\cos^2\theta}{\Sigma}\,d\phi\wedge d\psi\,,
    \end{align}
\end{subequations}
and one can check that these solve the BPS equations exactly. 
It is interesting to note that the two-form $\gamma_1$ receives no $\bt^2$ correction. 
Ansatz quantities not listed above, such as $Z_2$, do not receive any corrections, while $Z_A$ and $\tA$ can be trivially read off from \eqref{eq:PertAnsatz}.

\begin{figure}[t]
    \centering
    \includegraphics[width=\textwidth]{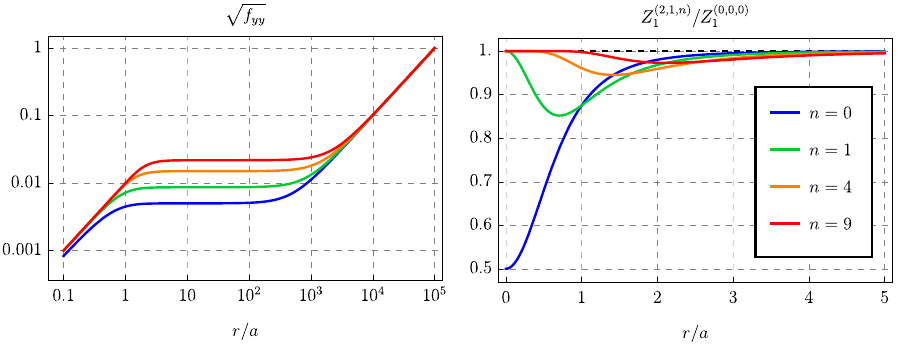}
    \caption{On the left, we plot $\sqrt{f_{yy}}$ as a function of the radial coordinate for various values of $n$ (the legend applies for both plots). This serves as a measure of the size of the $S_y^1$-circle. We see three distinct regions: An initial linear increase, signaling the AdS$_3$ cap. An intermediate constant region, denoting the finite AdS$_2\times S_y^1$ throat, which finally opens up into the asymptotic AdS$_3$ region. On the right, we plot the ratio between $Z_1^{(2,1,n)}$ and $Z_1^{(0,0,0)} =Q_1/\Sigma$. We see that only the $n=0$ solution starts at a value below 1, as can be seen in \eqref{eq:NearST21n}. In all others the minimum of the ``bump'' is located at a finite value of $r$. In these plots, we have taken $Q_1 = Q_5 = 10^{10}$, $R_y = 10^7$, and $a=1$. In the left plot $\theta = \pi/4$, while as can be seen from \eqref{eq:RatioZ1}, the right plot holds for all $\theta$. }
    \label{fig:21nFig}
\end{figure}

The momentum and angular momenta can be read off as in the general case. 
Perhaps the most interesting property of this family is the very simple correction in $Z_1$.
Indeed, one notices that the entire $\theta$ dependence in $Z_1^{(2,1,n)}$ comes from the overall $\Sigma^{-1}$ factor
\begin{equation}
\label{eq:RatioZ1}
    Z_1^{(2,1,n)} = \frac{Q_1}{\Sigma}\left[1- \frac{b^2\,R_y^2}{2\,Q_1\,Q_5}\,\frac{a^4}{\left(a^2+r^2\right)^2}\,\Delta_{0,0,2n}\right].
\end{equation}
In fact, the prefactor of the second term in the square bracket can be identified, through \eqref{eq:Translation}, with $(n+1)N_b/N$.
This explicitly shows how the presence of the additional strands in the full CFT state deforms the corresponding bulk geometry. 
The term in the square bracket is only a function of the radial coordinate, and we plot the ratio between $Z_1^{(2,1,n)}$ and the undeformed ansatz quantity $Z_1^{(0,0,0)}\equiv Q_1/\Sigma$ on the right plot in Figure~\ref{fig:21nFig}.
We notice that for all $n$, we find a ``bump'', whose minimum is located at some finite value of $r$, which is increasing with increasing value of $n$ (roughly like a square root, just as in \cite{Bena:2015bea, Bena:2017xbt}).
It is interesting to note that for $n=0$, the location of the bump is at $r=0$. This can also be seen by analysing the value of $Z_1$ at the location of the supertube using \eqref{eq:STLim}
\begin{align}
\label{eq:NearST21n}
    Z_1^{(2,1,n)} \xrightarrow[\lambda\to 0]{}\frac{1}{a^2\,\lambda^2}\left( Q_1 - \delta_{n,0}\frac{\tilde b^2}{2}\right)\,.
\end{align}
Since the value in the bracket corresponds to the local brane charge, one may wonder whether it is possible that this charge becomes negative. It turns out that the regularity condition \eqref{eq:ExplicitRegularity0} prohibits this: the charge at the supertube location is always  positive
\begin{align*}
     Q_1 - \delta_{n,0}\frac{\tilde b^2}{2} = \frac{1}{Q_5}\left( Q_1\,Q_5 - \delta_{n,0}\frac{\tilde b^2\,Q_5}{2}\right) = \frac{1}{Q_5}\left( a^2\,R_y^2+  \frac{\tilde b^2\,Q_5}{2}\right)\geq 0\,,
\end{align*}
where we have used that $x_{2,1,0}=2$.

The second quantity of interest is the size of the $S_y^1$ circle, which implicitly tells us about the local structure of the geometry. 
While the asymptotic region is AdS$_3 \times S^3$, we would like to know the structure of the interior region.
To that end, we first rewrite the six-dimensional metric as
\begin{align}
    ds^2_6 &=  f_{rr}\, dr^2 + f_{tt}\,\,\left(dt+ A^{(t)}\right)^2 + f_{yy}\,\left(dy+ A^{(y)}\right)^2\nonumber\\*
    &\quad +f_{\theta\theta}\,\,d\theta^2 +
    f_{\phi\phi}\,\left( d\phi + A^{(\phi)}\right)^2 + f_{\psi\psi}\,\left( d\psi + A^{(\psi)}\right)^2\,,
    	\label{eq:FiberedMetric}
\end{align}
where the $A^{(M)}$ are one-forms on  AdS$_3$. 
The size of the $y$-circle is proportional to $\sqrt{f_{yy}}$, which we plot on the left in Figure~\ref{fig:21nFig}.
We find that all the geometries change as AdS$_3\leftrightarrow$AdS$_2\times S_y^1\leftrightarrow$AdS$_3$ as the radial coordinate is varied. 
We thus see that the geometries develop a long, but finite throat, whose presence mimics the behaviour of an infinite throat for black holes.
However, unlike in black holes, these geometries smoothly cap off in another AdS$_3$ region, before ever developing a horizon. 
It can be shown that this property is present for solutions with arbitrary mode numbers $(k,m,n)$.


\section{Discussion}
\label{sec:Discussion}

We have presented the construction of a new class of vector superstrata and have proposed their dual states in the D1-D5 CFT. 
With this, we have completed the set of superstrata in which the momentum is carried by just a vector field while the base space remains unchanged. 
We have also presented a new technique to solve Laplace equations in spheroidal coordinates that appear in the constructions of microstate geometries and we hope that this may prove useful in future constructions of new geometric solutions.

There are several future directions which are interesting. 
Firstly, our proposal for the identifications of CFT states with the supergravity solutions is based upon some assumptions which may not be well founded. 
It would be crucial to perform some detailed precision tests to confirm our proposal \cite{Kanitscheider:2006zf, Kanitscheider:2007wq, Taylor:2007hs, Giusto:2015dfa, Bombini:2017sge, GarciaiTormo:2019ssf, Giusto:2019qig, Giusto:2020mup, Rawash:2021pik, Ceplak:2021wak}. 
Alternatively, we could alleviate some problems by coupling our solutions to flat space \cite{Bena:2017xbt,Ceplak:2018pws,Govaerts:2023xxv},  since the methods used to extract the charges are valid in this case.
Extending the geometries to flat space is also of broader interest, as it would enhance our understanding of microstates of asymptotically-flat black holes.
In addition, it would allow for the study of potential emission of Hawking radiation  \cite{Avery:2009tu}, which is not possible in asymptotically AdS solutions.

Next, in this paper we limited ourselves to single mode superstrata: geometries where the momentum is sourced by a vector field with a single non-trivial Fourier component. 
However, one can consider solutions with several components turned on. As shown in \cite{Bena:2017xbt} and \cite{Heidmann:2019zws, Heidmann:2019xrd}, one can find smooth, consistent solutions even in this case, however, regularity imposes non-trivial constraints between the parameters $b_{(k,m,n)}$ associated with each mode.
It would be interesting to see whether something similar also happens for superstrata based on vector fields.
In particular,  it was shown that the consistent set of regularity conditions  requires two types of ``tensor-mode'' superstrata. 
These two-types are roughly analogous to vector superstrata found in \cite{Ceplak:2022pep} and those constructed in this paper, meaning that our newly built geometries may be crucial in multi-mode vector superstrata.

What is more, since we now have control over superstrata based on tensor \emph{and} vector-field excitations, it may possible to construct ``mixed'' multi-mode superstrata. 
In other words,  there may exist solutions in which momentum is carried by both vector and tensor fields and in which these excitations are coupled. 
The BPS equations for minimal supergravity in six dimensions coupled to arbitrary numbers of vector and tensor multiplets are in general not upper-triangular \cite{Cano:2018wnq}. Nonetheless, it may be that the coupling constants that are inherited from the ten-dimensional D1-D5 system conspire to allow for a closed form analytic solution.
It would be interesting to explore this avenue further.


Recently, a lot of progress has been made by utilising a three-dimensional truncation linked to tensor-mode superstrata with quantum numbers $(1,m,n)$ \cite{ Ganchev:2021iwy, Ganchev:2021pgs, Ganchev:2022exf, Ganchev:2023sth, Houppe:2024hyj} .
In particular, using this setup one is able to construct non-BPS microstate geometries.
It would be interesting to see whether vector superstrata admit a similar truncation and if so, if one is able to obtain non-BPS microstate geometries based on vector momentum carriers.

One of the features of vector superstrata is that there exist a frame in ten dimensions in which these geometries are described purely in the NS-NS sector \cite{Ceplak:2022pep}.
As such, these geometries may present an interesting starting point to explore the landscape of stringy solutions by providing a suitable background for gauged Wess-Zumino-Witten (WZW) models \cite{Martinec:2018nco,Martinec:2019wzw, Martinec:2020gkv, Martinec:2022okx}.%
\footnote{ See also \cite{Chakraborty:2023kus} for worldsheet constructions of black hole microstates. }
However, these (three-charge) superstrata contain $v$-dependent momentum modes. It would be interesting to understand how this dependence arises from the gauging conditions in the WZW models. If constructed, these would present the first class of superstrata geometries constructed from the string world-sheet.
An alternative utilisation of this special purely NS-NS frame is to study the scrambling of infalling strings \cite{Martinec:2020cml, Ceplak:2021kgl, Guo:2024pvv}. 
In this case all the interaction between the background and the stringy probe can be, in principle, taken into account. 
This is especially important, since it was shown that tidal excitations of stringy modes are important to mimic the trapping behaviour of black holes and thus remedying the echo signals observed for point-like probes in microstate geometries \cite{Bena:2019azk}.


In \cite{Bena:2022fzf, Eckardt:2023nmn} a generalization of the building blocks of black hole microstructure were conjectured to exist based principles of local supersymmetry. 
In addition, \cite{Bena:2022wpl} motivates that the leading order contribution to the entropy should come from states which carry degrees of freedom living along at least one direction of $T^4$, further emphasising the role of the  internal manifold. 
In the ten-dimensional frame where the global charges of vector superstrata are D1-D5-P, the momentum-carrying vector field corresponds to a two-form field with one leg along a direction in the $T^4$. 
Vector superstrata thus present a modest attempt to include the detailed dynamics of the internal manifold. 
However, it would be interesting to see whether one could include non-trivial dependence on any of the internal directions \cite{Bena:2023rzm,Bena:2023fjx,Bena:2024qed}. Of course, this would take us out of the regime of validity of the six-dimensional supergravity used in this paper. Nevertheless, the solutions presented here may still present a suitable starting point for further investigation.

Finally, let us note that typically, including in our construction, microstate geometries are constructed by considering the internal manifold to be $T^4$. This is a judicious choice since the dynamics along these directions are taken to be trivial.
However, in certain regimes nontrivial dynamics on this may play an important role.
It would be interesting to study the differences between degrees of freedom of $T^4$ and $K3$, the other Ricci flat manifold which can normally be interchanged trivially.
%

 \section*{Acknowledgements}

We would like to thank Iosif Bena, Soumangsu Chakraborty, Bin Guo,  Anthony Houppe, Yixuan Li, and Nicholas P. Warner for interesting discussions.
This work was supported in part by the ERC Grant 787320 - QBH Structure.
In addition,  N\v{C} is supported by
the Science Foundation Ireland under the grant agreement 22/EPSRC/3832, while the work of SDH is supported  by KIAS Grant PG096301.

\appendix

\section{Further details of the free orbifold CFT}
\label{app:CFT}

In this appendix we provide some additional details about the D1-D5 CFT.
In particular, we establish the notation and present the picture in the Neveu-Schwarz  (NS) sector, where the dual states take the form of a coherent build up of chiral-primary operators.
We draw heavily from the reviews of the D1-D5 CFT \cite{David:2002wn, Avery:2010qw} while following the notation of \cite{Shigemori:2019orj, Shigemori:2020yuo}.

\subsection{Notation and conventions}

As in the main text, we focus on the D1-D5 CFT where on the internal manifold $\cM = T^4$. 
The symmetry group  is $SU(1,1|2)_L \times SU(1,1|2)_R$, which has several affine generators that we use to add momentum to the states
\begin{subequations}
    \begin{align}
    &\text{Left-moving generators}:\qquad   L_n\,, J^i\,, G_n^{\alpha A}\,,\\
    &\text{Right-moving generators}:\qquad  \Lt_n\,, \Jt^{\bar i}\,, \Gt_n^{\dot \alpha A}\,.
\end{align}
\end{subequations}
$L_n$ and $\Lt_n$ are the generators of the Virasoro symmetry, $SL(2, \mathbb{R})_L\times SL(2,\mathbb{R})_R$, $\alpha, \dot \alpha = \pm$ are doublet and $i, \bar i = 1,2,3$ the triplet indices of the R-symmetry group $SU(2)_L\times SU(2)_R$.
Finally $G_n^{\alpha A}$ and its right-moving counterpart denote the supersymmetry generators.
The index $A = 1,2$ denotes a doublet of the $SU(2)_B$ outer automorphism symmetry of the boundary superalgebra.
There is a custodial group $SU(2)_C$, and we use $\dot A= 1,2$ to denote its doublet indices. 
The last two groups combine to form the symmetry group of the $T^4$,  \mbox{$SU(2)_B\times SU(2)_C=SO(4)_I$}.
As such, an object with both $A$ and $\dot A$ indices can be interpreted as an $SO(4)_I$ vector%
\footnote{One can translate from one description to the other using \cite{Avery:2010qw} 
\begin{align*}
	X^{\dot A A} = \frac{1}{\sqrt 2}\, X^a\, (\sigma^a)^{\dot A A}\,,
\end{align*}	
where $\sigma^{a=1,2,3}$ are Pauli matrices and $\sigma^{a=4} = i\mathbb{I}_2$.}
which can be given a physical interpretation of a polarisation of the D1-brane inside the D5-brane world-volume. 
Then, when dimensionally reducing type IIB theory on $T^4$ to six dimensions, the $C_2$ field that contains the information about these D1-branes  have  legs along a direction of four-torus and give rise to vector fields in the lower-dimensional theory \cite{Ceplak:2022pep}.

\subsection{States in the NS-NS sector}
In the main text we have only focused on the states in the Ramond-Ramond sector of the theory, which are naturally associated with black holes in the bulk. 
However, the initial stages of our construction are somewhat more straightforward in the NS-NS sector, which is why we present some basic properties of this sector in this appendix.

The basic building blocks in this sector are 1/4-BPS anti-chiral primary states%
\footnote{One can equivalently use chiral-primary states. Here we follow the convention of the microstate geometries program.}
\cite{Shigemori:2019orj, Shigemori:2020yuo}
\begin{subequations}
	\label{eq:ACP}
	\begin{align}
		&\kett{ \alpha \dot \alpha}^{\rm NS}_k\,,  &&h^{\rm NS} = - j^{\rm NS} = \frac{k + \alpha}{2}\,,\qquad  && \hbn = - \jbn =  \frac{k + \dot \alpha}{2}&& \text{boson}\,,\label{eq:nsbos1}\\
		&\kett{ \alpha \dot A}^{\rm NS}_k\,,  &&h^{\rm NS} = - j^{\rm NS} = \frac{k + \alpha}{2}\,, \qquad && \hbn = - \jbn =  \frac{k }{2}\,, && \text{fermion}\label{eq:nsfer1}\\
		&\kett{ \dot A \dot \alpha}^{\rm NS}_k\,, &&h^{\rm NS} = - j^{\rm NS} = \frac{k }{2}\,,\qquad  && \hbn = - \jbn =  \frac{k + \dot \alpha}{2}&& \text{fermion},\label{eq:nsfer2}\\
		&\kett{ \dot A  \dot B }^{\rm NS}_k\,,  &&h^{\rm NS} = - j^{\rm NS} = \frac{k }{2}\,, \qquad && \hbn = - \jbn =  \frac{k }{2}&& \text{boson}\label{eq:nsbos2}.
	\end{align}
\end{subequations}
where $h^{\rm NS}$ and $j^{\rm NS}$ are the eigenvalues
under the action of $L_0$ and  $J_0^3$ respectively, and the barred quantities are their right-moving analogues.
As in the main text, $k = 1,2,\ldots, N$ denotes the length of the effective string (or more properly, the twist sector of the state) and  $\alpha, \dot \alpha = \pm$ should be interpreted as $\pm 1$. 
The spin of the state is given by  $s = j^{\rm NS} - \jb^{\rm NS}$. 
Note that for all states in the NS-NS sector we add a superscript ``NS'' to distinguish them from the R-R sector states which are described in the main text and do not have a superscript. 

The state $\kett{--}^{\rm NS}_1$  denotes the vacuum state, with $h^{\rm NS} = \hb^{\rm NS} = j^{\rm NS} =\jb^{\rm NS}=0$, while other anti-chiral primaries can be interpreted  as single-particle states and correspond to perturbations around global AdS$_3\times S^3$ in the bulk \cite{Deger:1998nm, deBoer:1998ip}.

\paragraph{CFT state related to vector fields in the NS-NS sector}

From a group-theoretic point of view, the anti-chiral primaries \eqref{eq:ACP} are the lowest-weight states in their respective multiplets. 
Their descendants are created by (repeated) action of the generators of the global symmetry of the CFT $L_{-1}$, $J_0^+$ and $G_{-\frac12}^{+A}$  and their right-moving counterparts.  
Acting with both left and right moving generators completely breaks supersymmetry, so we act only with the former and thus preserve 1/8 of the total supercharges of Type IIB supergravity.

In particular, we are interested in bosonic 1/8-BPS states, since they can be represented in the bosonic sector of the dual bulk supergravity theory. 
Bosonic descendants that have both an $A$ and $\dot A $ index are obtained by starting with a fermionic anti-chiral primary (either \eqref{eq:nsfer1} or \eqref{eq:nsfer2}) and acting on it with a supersymmetric generator $G_{-\frac12}^{+A}$.
This generates 16 new bosonic states, which are again the lowest weight states of their respective multiplet. 
All of them have already been described in Section 2 of \cite{Ceplak:2022pep}, but only 4 have been used to create microstate geometries. 
In this paper, we focus on the set
\begin{align}
	\label{eq:StartState}
	\kett{\psi}^{\rm NS}&\equiv G_{-\frac12}^{+A}\kett{- \dot A}_k^{\rm NS} &&h^{\rm NS} = \frac{k}{2}\,, \qquad j^{\rm NS} = -\frac{k-2}{2}\,,&& \hbn = - \jbn =  \frac{k }{2}\,,
\end{align}
which satisfy
\begin{align}
	\label{eq:StateProperty}
	J_0^-\,\kett{\psi}^{\rm NS}=  \Jb_0^-\,\kett{\psi}^{\rm NS} = L_1\,\kett{\psi}^{\rm NS} = \Lb_1\,\kett{\psi}^{\rm NS}
	= 0\,,
\end{align}
since they are the lowest weight states in their $SU(2)_L\times SU(2)_R$ and $SL(2,\mathbb{R})_L\times SL(2,\mathbb{R})_R$ multiplets.
Supersymmetric descendant states can be created by acting on this state with $L_{+1}$ and $J_0^{+}$, which yields the state
\begin{align}
	\label{eq:NSstate}
	\kett{k,m,n;\dot A, A}^{\rm NS} \equiv \left(L_{-1}\right)^{n} \left(J_0^{+}\right)^{m-1}G_{-\frac12}^{+A}\kett{- \dot A}_k^{\rm NS}\,,
\end{align}
whose eigenvalues are given by
\begin{align}
	\label{eq:chargesstate1}
	&&h^{\rm NS} = \frac{k}{2}+n\,, \qquad j^{\rm NS} = -\frac{k}{2}+m\,,&& \hbn = - \jbn =  \frac{k }{2}\,. 
\end{align}
In the above $k=2,3,\ldots$, $m =  1, \ldots, k$, and 
 $n =0, 1, 2, \ldots$.%
\footnote{The state \eqref{eq:StartState} does not exist for $k=1$, because $\left|\right.-\dot A\left.\right\rangle_1^{\rm NS}$ is an $SU(2)_L$ singlet and thus gets annihilated by $G_{-\frac12}^{+A}$.}

As already discussed in the main text, the full state in the D1-D5 CFT is comprised of several strands with different lengths, identified under permutations. 
In our case,  we consider $N_a$ copies of the vacuum state $\kett{--}_1^{\rm NS}$ and $N_b$ copies of states \eqref{eq:NSstate}
\begin{align}
	\label{eq:FullStateNS}
	\left(\kett{--}_1^{\rm NS}\right)^{N_a}\, \left(\kett{k,m,n; \dot A,A}^{\rm NS} \right)^{N_b}\,,
\end{align}
subject to the total length constraint
\begin{align}
	\label{eq:AStrandBudget}
	N = N_a + k \, N_b\,.
\end{align}
When $N_b=0$,  the full state is that of the global vacuum and is holographically dual to global AdS$_3 \times S^3$.
For $N_b=1$, the eigenvalues of \eqref{eq:FullStateNS} are given by  \eqref{eq:chargesstate1} and, on the bulk side,  corresponds to an 1/8-BPS excitation around global AdS$_3 \times S^3$.

States in the CFT that are associated with microstate geometries lie in the Ramond  sector. 
To go from the NS to R sector, we use the spectral flow which maps
\cite{Avery:2010qw, Shigemori:2020yuo}%
\footnote{Setting $\eta= \frac12$ maps an anti-chiral primary to a Ramond ground state, and vice-versa with $\eta = - \frac12$.}
\begin{align}
	\label{eq:SpectralFlowCFT}
	h' ~=~ h + 2\,\eta\,j + k \, \eta^2\,, \qquad j'~=~  j + k\, \eta\,.
\end{align}
Under this transformation the vacuum state maps to $\kett{++}_1$ state that comprises the round supertube \eqref{eq:Supertubestate}
\begin{align}
	\label{eq:vacuumtrans}
	&\kett{--}_1^{\rm NS} \mapsto \kett{++}_1\,, && h = \hb = \frac14\,, && j = \jb = \frac{1}{2}\,,
\end{align}
while the state \eqref{eq:NSstate} becomes 
\begin{align}
	\label{eq:ARstate}
	\kett{k,m,n; \dot A,A}^{\rm NS}\mapsto  \left(L_{-1}-J_{-1}^3\right)^{n}\, \left(J_{-1}^+\right)^{m-1} \, G_{-1}^{+A}\,\kett{+\dot A}_k \equiv \kett{k,m,n; \dot A, A}\,,
\end{align}
which we have used in the main part of the text \eqref{eq:Rstate}.
Thus, when combined, the spectral flow of the full state \eqref{eq:FullStateNS} is the state \eqref{eq:FullStateR} that we proposed as the CFT dual of the family of superstrata that we have constructed.%
\footnote{As in the main text, we are ignoring the subtleties with coherent sums and invariance under the permutation group $S_N$.}

\section{Details of the supergravity calculation}
\label{app:SolLay}

In the main text, we have presented a systematic way to solve the layered BPS equations  starting from the perturbation \eqref{eq:PertRkmn}. 
In doing so, we have omitted some details, which we present in this appendix.
In addition, we show how to generate the aforementioned perturbation using the CFT data presented in Appendix~\ref{app:CFT}.
For concreteness, throughout this appendix, we focus on a single mode excitation, with $\bt_{(k,m,n)} \equiv \bt$.

\subsection{Generating the perturbation}

To generate the perturbation, it is more convenient to work in the NS sector of the theory.
The vacuum state, \eqref{eq:FullStateNS} with $N_b=0$, is holographically dual to global AdS$_3\times S^3$ 
\begin{align}
\label{eq:GlobalAdS}
	ds^2 =- \;\! \frac{r^2+a^2}{a^2 R_y^2}dt^2 +\frac{r^2}{a^2 R_y^2}dy^2 
	+\frac{dr^2}{r^2+a^2}
	+d\theta^2+\sin ^2\theta\, d\phit^2+\cos ^2\theta \,d\psit^2\,,
\end{align}
with radii $R_{\rm AdS_3}^2 = R_{\rm S^3}^2 = \sqrt{Q_1\,Q_5}$.
There is an additional three-form $G$, given by the sum of the volume forms of the AdS$_3$ and $S^3$, which supports the non-trivial curvature.

Let us now focus on the state \eqref{eq:FullStateNS} with $N_b =1 \ll N_a$, which is dual to a supersymmetric perturbation around global AdS$_3\times S^3$.
The state \eqref{eq:NSstate} has two free indices $A$ and $\dot A$ combining into the symmetry of the $T^4$, $SU(2)_B \times SU(2)_C \simeq SO(4)_I$, hence we assume that the dual bulk excitation has a  leg in the four-torus. 
When viewed from the six-dimensional description, this generates a perturbation with a vector-field, or equivalently, a one-form.
Following \cite{Deger:1998nm}, we use the isometries of the metric \eqref{eq:GlobalAdS} to make an ansatz for the one-form
\begin{align}
	\label{eq:BPertAns}
	A^{\rm NS} = \bt\, \big[f(r) \, A_{\rm S^3}(\theta) + g(\theta) \, A_{\rm AdS_3}(r)\big]\, e^{i\left( m_1 \phit + m_2\psit + \frac{n_1}{R_y}\, t + \frac{n_2}{R_y}\,y\right)}\,,
\end{align} 
where   $f(r)$ and $g(\theta)$ are arbitrary functions, $ A_{\rm AdS_3}(r)$ and $ A_{\rm S^3}(\theta)$ are one-forms on AdS$_3$ and $S^3$ respectively, and we have introduced free parameters $n_{1,2}$ and $m_{1,2}$. 
Finally, $\bt \ll a$ is a length parameter measuring the strength of the perturbation and is related to $N_b/N$. 

To relate this perturbation to the state \eqref{eq:FullStateNS}, we impose that it has the same properties under the symmetry algebra of the background.
This is possible because the representation of the symmetry generators is known for global AdS$_3 \times S^3$ \cite{Maldacena:1998bw}
\begin{subequations}
	\begin{gather}
		\begin{split}
			L_0&={i R_y \over 2}(\partial_t+\partial_y),\\
			L_{\pm 1}
			&=ie^{\pm {i\over R_y}(t+y)}
			\biggl[
			-{R_y\over 2}\biggl({r\over \sqrt{r^2+a^2}}\partial_t+{\sqrt{r^2+a^2}\over r}\partial_y\biggr)
			\pm {i\over 2}\sqrt{r^2+a^2}\,\partial_r
			\biggr],\\
			\Lb_0 &= \frac{i R_y }{2}\left( \partial_t - \partial_y\right)\,,\\
			\Lb_{\pm1}&=  i e^{\pm {i\over R_y}(t-y) } \biggl[
			-{R_y\over 2}\biggl({r\over \sqrt{r^2+a^2}}\partial_t-{\sqrt{r^2+a^2}\over r}\partial_y\biggr)
			\pm {i\over 2}\sqrt{r^2+a^2}\,\partial_r
			\biggr],
		\end{split} 
	\end{gather}
\end{subequations}
and \cite{Giusto:2013bda}
\begin{subequations}\\
	\label{eq:SU(2)gen}
	\begin{align}
		&J_0^3=-{i\over 2}(\partial_{\phit}+\partial_{\psit}),&&
		J_0^\pm ={1\over 2}e^{\pm i(\phit+\psit)}
		(\pm  \partial_\theta+i\cot\theta\, \partial_{\phit}-i\tan\theta\, \partial_{\psit})\,,\\
		&\Jb_0^3 = - \frac{i}{2}\left( \partial_\phit - \partial_\psit\right)\,,&& 
		\Jb^{\pm}_0 = \frac{1}{2}e^{\pm i (\phit - \psit)} \left( \mp \partial_\theta - i \cot\theta\, \partial_\phit - i \tan\theta \,\partial_\psit\right)\,.
	\end{align}
\end{subequations}
In particular, we want to impose  that the perturbation has the same eigenvalues as \eqref{eq:StartState} and obeys \eqref{eq:StateProperty}
\begin{subequations}
	\label{eq:PertCons}
	\begin{gather}
	 J_0^3 \, A^{\rm NS}	= -\frac{k-2}{2}\,A^{\rm NS}\,,\qquad   L_0\, A^{\rm NS} = \Lb_0 \, A^{\rm NS} =  - \Jb^3_0 \, A^{\rm NS} = \frac{k}{2}\,A^{\rm NS}\,,\\
		J_0^- \, A^{\rm NS} =  \Jb_0^- \, A^{\rm NS} =  \Lb_1 \,A^{\rm NS} = 0\,,
	\end{gather}
\end{subequations}
which are solved by%
    \footnote{There is an additional degree of freedom
    set to 0 by imposing that the corresponding solution in the Ramond sector has no $du$ component, which is imposed by the supersymmetry ansatz \eqref{eq:Potentials}.}
\begin{align}
	A^{\rm NS} = -\bt \, \Delta_{k,1,0}\,e^{-i \left(  k\left(\phit+\frac{t}{R_y}\right)- \phit -\psit\right)}\left[\frac{i\,d\theta}{\sin\theta\cos\theta}+ d\phit + d\psit\right]\,, 
\end{align}
where
\begin{align}
	\Delta_{k,m,n} &\equiv
	\left(\frac{a}{\sqrt{r^2+a^2}}\right)^k
	\left(\frac{r}{\sqrt{r^2+a^2}}\right)^n 
	\cos^{m}\theta \, \sin^{k-m}\theta \,.
\end{align}
Following \eqref{eq:NSstate}, we obtain the gravity duals of the state with arbitrary numbers $(k,m,n)$ through the action of $L_{-1}$ and $J_0^+$ generators, which yields
\begin{align}
\label{eq:NSPert}
	A_{k,m,n}^{\rm NS} = -\bt \, \Delta_{k,m,n}\,e^{-i \left(n \frac{t + y}{R_y} + k\left(\phit+\frac{t}{R_y}\right)- m (\phit + \psit)\right)}\left[\frac{i\,d\theta}{\sin\theta\cos\theta}+ d\phit + d\psit\right]\,. 
\end{align}

This is the perturbation dual to the state \eqref{eq:FullStateNS} in the NS sector. 
In the main text we used the perturbation in the Ramond sector, which is obtained by using the change of coordinates corresponding to the  spectral flow 
\begin{align}
	\label{eq:SpectralFlow}
	\phit = \phi - \frac{t}{R_y}, \qquad \psit = \psi - \frac{y}{R_y}\,.
\end{align}
The metric \eqref{eq:GlobalAdS} becomes that of the maximally spinning supertube \eqref{eq:Supertube}, as expected from \eqref{eq:vacuumtrans}, while \eqref{eq:NSPert} is now
\begin{align}
	A_{k,m,n} = \bt\,e^{- i v_{k,m,n}}\, \Delta_{k,m,n}\, \left[\frac{\sqrt{2}}{R_y}\, dv -
	\frac{i\,d\theta}{\sin\theta\cos\theta}- d\phi - d\psi\right] \,, 
\end{align}
and we used 
\begin{align}
	\label{eqBvhatPhase}
	\hat{v}_{k,m,n} &\equiv (m+n) \frac{\sqrt{2}\,v}{R_y} + (k-m)\phi - m\psi \,.
\end{align}
Taking the real part of the above perturbation yields 
\begin{align}
	\label{eq:BPertRkmn}
	A_{k,m,n} = \bt\, \Delta_{k,m,n}\, \left[\left( \frac{\sqrt{2}}{R_y}\, dv-d\phi - d\psi\right) \,\cos{\hat v_{k,m,n}}-  \frac{d\theta}{\sin\theta\cos\theta}\, \sin{\hat v_{k,m,n}} \right] \,,
\end{align}
which is the starting point of our calculation in the main text.

This perturbation can be put into the supersymmetric ansatz \eqref{eq:Potentials}, which allows us to read off
\begin{subequations}
	\label{eq:BPertAnsatz}
	\begin{align}
		Z_A^{(k,m,n)} &\equiv \frac{\sqrt{2}\,\bt\,Q_5}{R_y}\, \frac{\Delta_{k,m,n}}{\Sigma}\,\cos{\hat v_{k,m,n}}\,,\\
		\tA^{(k,m,n)} &\equiv  \bt\, \Delta_{k,m,n}\, \left[ \frac{d\theta}{\sin\theta\cos\theta}\, \sin{\hat v_{k,m,n}}  + \frac{(a^2 + r^2) \,d\phi + r^2\, d\psi}{\Sigma}\,\cos{\hat v_{k,m,n}} \right]\,.
	\end{align}
\end{subequations}
 Using \eqref{eq:U1Comp}, we can show the components of the two-form field strength, $F$, are given by
\begin{subequations}
	\label{eq:OmFtF}
	\begin{align}
		\omega_F^{(k,m,n)} &= \frac{\sqrt2\,\bt}{R_y}\,\Delta_{k,m,n}\Bigg\{-\Bigg[\frac{n\,a^2- k\,r^2}{r\left(a^2 + r^2\right)}\,dr+ \left(k \cot\theta + \frac{n}{\sin\theta\, \cos\theta}\right)\,d\theta\Bigg 
  ]\cos\vh_{k,m,n}\nonumber\\
  & \hspace{5cm} + \Bigg[(k+n)\, d\theta + n \, d\psi\Bigg]\sin\vh_{k,m,n}\Bigg\}
		\,,\\
		\tF^{(k,m,n)} &= -\bt\,\Delta_{k,m,n}\Bigg[\left(\frac{n\,a^2- k\,r^2}{r\,\sin\theta}\right)\Omega_1\,\sin\vh_{k,m,n}+\frac{n\,a^2- k\,r^2}{\Sigma}\left(\Omega_2 + \Omega_3\right)\, \cos\vh_{k,m,n} \Bigg]
 \,,
	\end{align}
\end{subequations}
where we have  introduced a basis of self-dual two-forms  $\mathbb{R}^4$
\begin{subequations}
	\label{eq:SDBas}
	\begin{align}
		\Omega_{1} &\equiv \frac{dr\wedge d\theta}{(r^2+a^2)\cos\theta} + \frac{r\sin\theta}{\Sigma} d\phi\wedge d\psi\,,\\
		\Omega_{2} &\equiv  \frac{r}{r^2+a^2} dr\wedge d\psi + \tan\theta\, d\theta\wedge d\phi\,,\\
		\Omega_{3} &\equiv \frac{dr\wedge d\phi}{r} - \cot\theta\, d\theta\wedge d\psi\,,
	\end{align}
\end{subequations}
which explicitly show that $\tF$ is self-dual on $\IR^4$.
Indeed, one can show that \eqref{eq:OmFtF} solve the second-layer BPS  equations \eqref{eq:Lay2}.

\subsection{Solving the third layer}

When $\bt\sim a$, which corresponds to $N_b \sim N_a$ on the CFT side, then \eqref{eq:BPertRkmn} significantly deforms the background and one needs to solve the BPS equations to all order in $\bt/a$.
Thankfully, the supertube background and \eqref{eq:BPertRkmn} solve the first and second layer equations to all orders in $\bt$, so we only need to solve the last two layers of BPS equations. 
Due to the upper-triangular structure, one can go in order and first consider the third layer equations \eqref{eq:Lay3} that determine  $\Theta^2$ and $Z_1$.

One can check that the source terms $\tF \wedge \tF$ and $\omega_F\wedge \tF$ are $v$-independent, from which we assume that $Z_1$ and  $\Theta^2$ are also  $v$-independent.
This simplifies their decomposition \eqref{eq:VectorInfluence} to 
\begin{subequations}
	\label{eq:BTheta2vindAns}
	\begin{gather}
		\Theta^2 ~=~ d_4a_2 + \tA \wedge \omega_F + \frac{Z_A}{Z_2}\, \tF  \\
		*_4 d_4 Z_1~=~ d_4 \gamma_1 - a_2 \wedge d_4\beta- \tF\wedge \tA\,,\label{eq:seceq}
	\end{gather}
\end{subequations}
which are then inserted into the BPS equations.
Only the self-duality condition for $\Theta^2$ is not trivially satisfied and reduces to 
\begin{gather}
	\label{eq:a2eq}
	d_4a_2 -  *d_4 a_2 = \frac{\sqrt{2}\,\bt^2}{R_y}\, \Delta_{2k,2m,2n}\left(  \frac{(k+n)}{\sin^2\theta}\,\oO_2-\frac{n}{\cos^2\theta}\oO_3\right) \,,
\end{gather}
with the right-hand side expressed in terms of a basis of anti-self dual two-forms on $\IR^4$
\begin{subequations}
	\label{eq:asdbasis}
	\begin{align}
		\oO_1 &= \frac{dr\wedge d\theta}{(a^2 + r^2) \cos\theta} - \frac{r \sin\theta}{\Sigma}\, d\phi \wedge d\psi\,,\\
		\oO_2 &= \frac{r}{a^2 + r^2}\, dr\wedge d\psi - \tan\theta d\theta \wedge d\phi\,,\\
		\oO_3 &= \frac{dr \wedge d\phi}{r}+ \cot\theta \, d\theta \wedge d\psi\,.
	\end{align}
\end{subequations}
Since there is no term in \eqref{eq:a2eq} that is proportional to $\oO_1$, which contains $dr\wedge d\theta$, one can make an ansatz
\begin{gather}
	\label{eq:BTheta2ans}
	a_2 =\frac{\bt^2}{\sqrt2\,R_y}\biggr[ f_1(r,\theta)\left( d\phi + d\psi \right) + f_2 (r,\theta) \left( d\phi - d\psi\right)\biggr]\,.
\end{gather}
Inserting this into \eqref{eq:a2eq} yields two differential equations for $f_1(r,\theta)$ and $f_2(r,\theta)$,
\begin{subequations}
	\label{eq:oocomp}
	\begin{gather}
		- \frac{\cot}2\theta\left(\pd_\theta f_1 + \pd_\theta f_2\right)+ \frac{a^2 + r^2}{2\,r}\left(\pd_r f_1 - \pd_r f_2\right)=  \frac{k+n}{\sin^2\theta}\,\Delta_{2k,2m,2n}\,,\\
         + \frac{\tan\theta}{2}\left(\pd_\theta f_1 - \pd_\theta f_2\right)+ r\,\left(\pd_r f_1 + \pd_r f_2\right)=- \frac{n}{\cos^2\theta}\, \Delta_{2k,2m,2n}\,,
	\end{gather}
\end{subequations}
which can be used to express the derivatives of $f_1$, purely in terms of $f_2$
\begin{subequations}
    \label{eq:f1eqB}
    \begin{align}
        \pd_r f_1 &= \frac{2\,k\,r}{r^2 + a^2 \,\sin^2\theta}\, \Delta_{2k,2m,2n} + \frac{2\,r\,\sin\theta\,\cos\theta}{r^2 + a^2 \,\sin^2\theta}\,\pd_\theta f_2 + \left(1- \frac{2\,r^2\,\cos^2\theta}{r^2 + a^2\,\sin^2\theta}\right)\,\pd_r\,f_2\,,\label{eq:f1reqB}\\
        \pd_\theta f_1 &= \left(1- \frac{2\,r^2\,\cos^2\theta}{r^2 + a^2\,\sin^2\theta}\right)\,\pd_\theta f_2 - 2\,\frac{\left(a^2 +r^2\right)\,r\,\sin\theta\,\cos\theta}{r^2 + a^2\,\sin^2\theta}\,\pd_r f_2\nonumber\\*
        &\quad -\frac{2 \tan\theta}{r^2 + a^2\,\sin^2\theta}\left(k\,r^2\,\cot^2\theta + n\,\frac{r^2 + a^2\,\sin^2\theta}{\sin^2\theta}\right)\,\Delta_{2k,2m,2n}\,\label{eq:f1theqB}.
    \end{align}
\end{subequations}
One can then rearrange the equations in such a way that one is left with a Laplace equation determining $f_2$
\begin{equation}
	\label{eq:Lap1B}
	\cLh f_2 = -\frac{2\,a^2}{\Sigma(a^2 + r^2)\, \cos^2\theta}\,\Big(n^2\, \Delta_{2k,2m,2n-2} + (k+n)^2\, \Delta_{2k,2m+2,2n}  -k\,(m+n)\, \Delta_{2k-2, 2m, 2n}\Big)\,,
\end{equation}
where $\cLh$ is the scalar Laplace operator on the 4-dimensional base space
\begin{equation}
	\label{eq:BLapDef}
	\cLh f(r,\theta) \equiv - *_4 d_4 *_4 d_4\,f =\frac{1}{r\, \Sigma}\,\pd_r \left(r\left(a^2 + r^2\right)\pd_r f\right)+ \frac{1}{\Sigma\, \cos\theta\, \sin\theta}\,\pd_\theta\left(\cos\theta\, \sin\theta\pd_\theta f\right)\,.
\end{equation}
Differential equation of the form
\begin{align}\label{solvableeq}
	\widehat{\cL}F_{2k,2m,2n}={\Delta_{2k,2m,2n}\over (r^2+a^2)\cos^2\theta\,\, \Sigma} 
\end{align}
are solved by \cite{Bena:2015bea,Bena:2017xbt}
\begin{equation} \label{eq:BFfun}
	F_{2k,2m,2n}=-\!\sum^{j_1+j_2+j_3\le k+n-1}_{j_1,j_2,j_3=0}\!\!{j_1+j_2+j_3 \choose j_1,j_2,j_3}\frac{{k+n-j_1-j_2-j_3-1 \choose k-m-j_1,m-j_2-1,n-j_3}^2}{{k+n-1 \choose k-m,m-1,n}^2}
	\frac{\Delta_{2(k-j_1-j_2-1),2(m-j_2-1),2(n-j_3)}}{4(k+n)^2(a^2 + r^2)}\,,
\end{equation} 
with
\begin{equation} 
	{j_1+j_2+j_3 \choose j_1,j_2,j_3}\equiv \frac{(j_1+j_2+j_3)!}{j_1!\, j_2!\, j_3!}\,.
\end{equation} 
This gives
\begin{equation}
	\label{eq:Bf2sol}
	f_2^{(k,m,n)}(r,\theta) = -2 \,a^2\Big[(k+n)^2\, F_{2k,2m+2,2n} + n^2\, F_{2k,2m,2n-2}   - k\,(m+n)\, F_{2k-2, 2m, 2n} \Big]\,.
\end{equation}
One is able to find a complicated closed-form expression for $f_1$ in terms of linear combinations of hypergeometric functions by inserting \eqref{eq:Bf2sol} into \eqref{eq:f1eqB} and integrating. However, in practice it is much easier to determine  $f_1$  case-by-case for fixed $(k,m,n)$ once $f_2$ is known. 
With this, $a_2$ is determined and it is easy to check that it solves the third layer BPS equations \eqref{eq:Lay3}. 

It is curious  that the BPS equations in this layer (and in fact in the fourth layer as well) can be solved without explicitly calculating  $Z_1$ -- it is enough to now $*_4 d_4 Z_1$ through \eqref{eq:Theta2vindAns}.
But the value of $Z_1$ is still needed when analysing the properties of the solutions, since its near-supertube and large $r$ behaviour determine the regularity and asymptotic behaviour of the six-dimensional metric, respectively. 
To find $Z_1$, we apply $-*_4 d_4$ to \eqref{eq:seceq}
\begin{align}
	\label{eq:BZ1Lap}
	\cLh Z_1 &=  \frac{4\,a^2\, \bt^2}{\Sigma^3\,(r^2+ a^2
			\sin^2\theta)}\left(\,a^2 \,\sin\theta\,\cos\theta \,\pd_{\theta} f_2 -  r(a^2 + r^2) \pd_r f_2 \right)\nonumber\\
		&\quad +\frac{4 \,a^4\,\bt^2\,k}{\Sigma^3\,(r^2+ a^2
			\sin^2\theta)}\,\Delta_{2k,2m,2n} - \frac{2\,\bt^2}{\Sigma^2}\, \frac{\Delta_{2k, 2m, 2n-2}}{ \cos^2\theta\,\sin^2\theta}\,\left(\frac{n\,a^2- k\,r^2}{a^2+r^2}\right)^2\,.
\end{align}
The most singular source term scales as $\Sigma^{-3}$ near the location of the supertube, where $\Sigma\rightarrow 0$, which indicates that at order $\bt^2$ there is a contribution to $Z_1$ that scales as $\Sigma^{-1}$ near the singularity.
How to find solutions to this equations is discussed in Section~\ref{sec:Bootstrapy}.

\subsection{Solving the fourth layer}

Inserting the solutions of the previous three layers into the right-hand sides of the differential equations in the fourth layer shows that only one term is non-vanishing and is, in fact, $v$-independent.
\begin{align}
	Z_2 \, \omega_F^2 = \frac{2 \,\bt^2\, Q_5}{R_y^2}\, \frac{1}{{\Sigma\,(a^2 + r^2)\, \cos^2\theta}}\left[n^2\, \Delta_{2k,2m,2n-2} + (k+n)^2\, \Delta_{2k,2m+2,2n}\right]\,.
\end{align}
Therefore, we again assume that neither $\omega$ nor $\cft$ are $v$-dependent. In this case the fourth layer BPS equations decouple. 
The equation that determines $\cft$ is
\begin{align}
	\cLh\,\cft =\frac{4 \,\bt^2\, Q_5}{R_y^2}\, \frac{1}{{\Sigma\,(a^2 + r^2)\, \cos^2\theta}}\left[(k+n)^2\,\Delta_{2k,2m+2,2n} + n^2\, \Delta_{2k,2m,2n-2}\right]\,,
\end{align}
and is solved by
\begin{align}
	\label{eq:BcFsol}
	\cft^{(k,m,n)} =  \frac{4\, \bt^2\, Q_5}{R_y^2}\left[(k+n)^2\,F_{2k,2m+2,2n} + n^2\, F_{2k,2m,2n-2}\right]\,.
\end{align}
The equation that determines $\omega$ is
\begin{align}
	\label{eq:Omegaeqkmn}
	d_4 \omega + *_4 d_4\omega &= Z_2 \, \Theta^2 - \cft \, d_4\beta = \frac{\sqrt{2}\,\bt^2\,Q_5}{R_y}\left(g_{2}\,\Omega_2 + g_3\,\Omega_3 \right)\,,
\end{align}
where we expanded in the self-dual basis \eqref{eq:SDBas} and
\begin{subequations}
	\begin{align}
		g_2 &\equiv \frac{k\,\left(a^2 + r^2\right)\Delta_{2k,2m,2n}}{\Sigma\left(r^2 + a^2\,\sin^2\theta\right)} + \frac{(a^2 + r^2) \,\sin\theta\,\cos\theta}{\Sigma\left(r^2 + a^2\,\sin^2\theta\right)}\,\pd_\theta f_2(r,\theta)- \frac{(a^2 + r^2) \,r\,\cos^2\theta}{\Sigma\left(r^2 + a^2\,\sin^2\theta\right)}\,\pd_r f_2(r,\theta)\nonumber\\*
		& \quad - \frac{4\,a^2\,(a^2 + r^2) \,\cos^2\theta}{\Sigma^2}\,f(r,\theta) \,,\\
		g_3 &\equiv \frac{k\,r^2\,\Delta_{2k,2m,2n}}{\Sigma\left(r^2 + a^2\,\sin^2\theta\right)} + \frac{r^2\,\sin\theta\,\cos\theta}{\Sigma\left(r^2 + a^2\,\sin^2\theta\right)}\,\pd_\theta f_2(r,\theta)+ \frac{(a^2 + r^2) \,r\,\sin^2\theta}{\Sigma\left(r^2 + a^2\,\sin^2\theta\right)}\,\pd_r f_2(r,\theta)\nonumber\\*
		& \quad + \frac{4\,a^2\,r^2 \,\sin^2\theta}{\Sigma^2}\,f(r,\theta) \,, 
	\end{align}
\end{subequations}
and we defined
\begin{align}
    \cft \equiv \frac{4\,\bt^2\,Q_5}{R_y^2}\,f(r,\theta)\,.
\end{align}
The lack of an $\Omega_1$ term on the right-hand side of \eqref{eq:Omegaeqkmn} leads to an ansatz
\begin{align}
	\label{eq:BOmegaBAns}
	\omega_{\bt} = \frac{\sqrt2\,Q_5\,\bt^2}{R_y} \left[\mu(r,\theta)\, \left( d\phi + d\psi \right) + \nu (r,\theta) \left( d\psi - d\phi\right)\right]\,,
\end{align}
which, when inserted into \eqref{eq:Omegaeqkmn}, yields two independent equations that can be used to derive
\begin{subequations}
	\label{eq:nuEq}
	\begin{align}
		\pd_r\nu &= \frac{k\,r\,\left((a^2 + r^2) - (a^2 + 2\,r^2)\, \cos^2\theta\right) }{\Sigma\,\left(r^2 + a^2\,\sin^2\theta\right)^2}\Delta_{2k,2m,2n} -\frac{2\,r\,\sin\theta\,\cos\theta}{r^2 + a^2\,\sin^2\theta}\,\pd_\theta\mu\nonumber\\* 
        &\quad - \frac{\left(a^2 + r^2 - r^2\,\cot^2\theta\right)\,\sin^2\theta}{r^2 + a^2\,\sin^2\theta}\,\pd_r\mu
        - \frac{2\,r^2\,(a^2 + r^2)\, \sin^2\theta\,\cos^2\theta }{\Sigma\,\left(r^2 + a^2\,\sin^2\theta\right)^2}\,\pd_r  
		f_2\nonumber\\* 
		&\quad+ \frac{r\,\left((a^2 + r^2) - (a^2 + 2\,r^2)\, \cos^2\theta\right)\,\sin\theta\,\cos\theta }{\Sigma\,\left(r^2 + a^2\,\sin^2\theta\right)^2}\,\pd_\theta f_2 
		- \frac{4\,a^2\,r\,(a^2 + 2r^2)\,\sin^2\theta\,\cos^2\theta }{\Sigma^2\,\left(r^2 + a^2\,\sin^2\theta\right)}\,f\,, \label{eq:nuEqr}
		\\ 
		\pd_\theta \nu &=  -\frac{2\,k\,r^2\,\left(a^2 +r^2\right)\,\sin\theta\,\cos\theta }{\Sigma\,\left(r^2 + a^2\,\sin^2\theta\right)^2}\Delta_{2k,2m,2n} 
		+ \frac{r^2 \,\cos2\theta- a^2\,\sin^2\theta}{r^2 + a^2\,\sin^2}\,\pd_\theta\mu  \nonumber\\
		& \quad 
		+ \frac{2\,r \,(a^2 + r^2) \,\sin\theta\,\cos\theta}{r^2 + a^2\,\sin^2}\,\pd_r\mu 
		- \frac{2\,r^2\,(a^2 + r^2)\, \sin^2\theta\,\cos^2\theta }{\Sigma\,\left(r^2 + a^2\,\sin^2\theta\right)^2}\,\pd_\theta 
		f_2\nonumber\\* 
		& \quad
		- \frac{r\,\left(a^2 +r^2\right)\left((a^2 + r^2)- (a^2 + 2r^2)\cos^2\theta\right)\,\sin\theta\,\cos\theta }{\Sigma\,\left(r^2 + a^2\,\sin^2\theta\right)^2}\,\pd_r 	f_2
		+ \frac{a^2\,r^2\,(a^2 + r^2)\,\sin4\theta}{\Sigma^2\,\left(r^2 + a^2\,\sin^2\theta\right)}\,f
		\,.\label{eq:nuEqth}
	\end{align}
\end{subequations}
One can then differentiate these expressions and after a bit of algebra  one is left with a single second-order differential equation for $\mu$ 
\begin{align}
	\label{eq:BmuEq}
	&\cLh\left(\mu + \frac{r^2+ a^2\,\sin^2\theta}{2\, \Sigma}\,f\right) =\frac{a^2}{4\,(r^2+ a^2\,\sin^2\theta)}\,\left[\cLh\left(\frac{f_2}{\Sigma}\right) - \frac{1}{\Sigma}\,\cLh f_2\right]\nonumber\\* 
    &+ \frac{n^2\,a^4}{2\,\Sigma\,\left(a^2+ r^2\right)^2}\,\Delta_{2k,2m,2n-2}- \frac{k\,(k+2n)\,a^2}{2\,\Sigma\,\left(a^2+ r^2\right)}\,\Delta_{2k,2m,2n}- \frac{k\,a^4}{\Sigma^2\,\left(r^2+a^2\,\sin^2\theta\right)}\,\Delta_{2k,2m,2n}\nonumber\\* 
    &+\frac{n^2\,a^2 -  (2\,k\,n+n^2)\,r^2}{2\,\Sigma\,\left(a^2+r^2\right)\,\cos^2\theta}\,\Delta_{2k,2m,2n-2} + \frac{(k+n)\left((k-n)\,r^2- 2\,n\,a^2\right)}{2\,\Sigma\,\left(a^2+r^2\right)\,\cos^2\theta}\,\Delta_{2k,2m+2,2n}\,,
\end{align}
There are no known closed-form solutions of this equation for all values of $(k,m,n)$. To that end, we introduce the method presented in Section~\ref{sec:Bootstrapy}, which in principle solves the equation for arbitrary values of the mode numbers. 
The last remaining step is then to determine $\nu$, which is done by integrating equations \eqref{eq:nuEq}, which again needs to be done on a case by case basis.

\section{Explicit example in detail (4,2,3)}
\label{app:example}
We record in this appendix expressions for the ansatz quantities for the mode numbers $k=4$, $m=2$, and $n=3$:%
\footnote{The choice of these numbers is relatively random. However, we chose $k=4$ and $m=2$ as this is the first example where we have a state in the $SU(2)_L$ multiplet that is neither the highest nor the lowest weight state and is thus simplified by some symmetry. $n=3$ is chosen arbitrarily.}
\begin{align}
\mu^{(4,2,3)}&={1\over2880\left(r^2+a^2\cos ^2\theta
   \right)\left(\frac{r^2}{a^2}+1\right)^7 }\bigg(-\frac{20 r^{12} \cos ^2\theta}{a^{12}}-\frac{20 r^{10} \cos ^4\theta}{a^{10}}\nonumber\\
   &-\frac{110 r^{10} \cos ^2\theta}{a^{10}}-\frac{564 r^8 \cos ^6\theta}{a^8}+\frac{726
   r^8 \cos ^4\theta}{a^8}-\frac{564 r^8 \cos ^2\theta}{a^8}+\frac{498 r^6 \cos ^6\theta}{a^6}\nonumber\\
   &-\frac{541 r^6 \cos ^4\theta}{a^6}-\frac{538 r^6 \cos ^2\theta
   }{a^6}+\frac{414 r^4 \cos ^6\theta}{a^4}-\frac{691 r^4 \cos ^4\theta}{a^4}-\frac{142 r^4 \cos ^2\theta}{a^4}\nonumber\\
   &+\frac{78 r^2 \cos ^6\theta}{a^2}-\frac{135 r^2 \cos
   ^4\theta}{a^2}-\frac{96 r^2 \cos ^2\theta}{a^2}+\frac{12 r^{14}}{a^{14}}+\frac{94 r^{12}}{a^{12}}+\frac{312 r^{10}}{a^{10}}+\frac{591 r^8}{a^8}\nonumber\\
   &+\frac{672 r^6}{a^6}+\frac{444
   r^4}{a^4}+\frac{156 r^2}{a^2}+6 \cos ^6\theta-11 \cos ^4\theta-18 \cos ^2\theta+23\bigg)
   \\
   \nonumber
   \\
   \nu^{(4,2,3)}&={1\over2880\left(r^2+a^2\cos ^2\theta\right)\left(\frac{r^2}{a^2}+1\right)^7}\bigg(\frac{24 r^{14} \cos ^2\theta}{a^{14}}-\frac{20 r^{12} \cos ^4\theta}{a^{12}}+\frac{188 r^{12} \cos ^2\theta}{a^{12}}\nonumber\\
   &-\frac{60 r^{10} \cos ^6(\theta
   )}{a^{10}}-\frac{40 r^{10} \cos ^4\theta}{a^{10}}+\frac{594 r^{10} \cos ^2\theta}{a^{10}}-\frac{360 r^8 \cos ^6\theta}{a^8}+\frac{44 r^8 \cos ^4\theta}{a^8}\nonumber\\
   &+\frac{1096
   r^8 \cos ^2\theta}{a^8}+\frac{258 r^6 \cos ^6\theta}{a^6}-\frac{285 r^6 \cos ^4\theta}{a^6}+\frac{790 r^6 \cos ^2\theta}{a^6}-\frac{210 r^4 \cos ^6\theta}{a^4}\nonumber\\ 
   &+\frac{469 r^4 \cos ^4\theta}{a^4}+\frac{210 r^4 \cos ^2\theta}{a^4}-\frac{54 r^2 \cos ^6\theta}{a^2}+\frac{109 r^2 \cos ^4\theta}{a^2}+\frac{104 r^2 \cos
   ^2\theta}{a^2}\nonumber\\
   &-\frac{12 r^{14}}{a^{14}}-\frac{94 r^{12}}{a^{12}}-\frac{312 r^{10}}{a^{10}}-\frac{591 r^8}{a^8}-\frac{672 r^6}{a^6}-\frac{444 r^4}{a^4}-\frac{156 r^2}{a^2}-6
   \cos ^6\theta\nonumber\\
   &+11 \cos ^4\theta +18 \cos ^2\theta-23\bigg)
   \\
   \nonumber
   \\
   f^{(4,2,3)}&=-{1\over2880 a^2 \left(\frac{r^2}{a^2}+1\right)^7}\bigg(\frac{10 r^{10} \cos ^2\theta}{a^{10}}-\frac{156 r^8 \cos ^4\theta}{a^8}+\frac{216 r^8 \cos ^2\theta}{a^8}\nonumber\\ 
   &-\frac{720 r^6 \cos ^6\theta}{a^6}+\frac{222 r^6 \cos
   ^4\theta}{a^6}+\frac{575 r^6 \cos ^2\theta}{a^6}-\frac{414 r^4 \cos ^4\theta}{a^4}+\frac{449 r^4 \cos ^2\theta}{a^4}\nonumber\\
   &-\frac{78 r^2 \cos ^4\theta}{a^2}+\frac{87 r^2
   \cos ^2\theta}{a^2}+\frac{30 r^{12}}{a^{12}}+\frac{190 r^{10}}{a^{10}}+\frac{479 r^8}{a^8}+\frac{616 r^6}{a^6}+\frac{428 r^4}{a^4}+\frac{154 r^2}{a^2}\nonumber\\ 
   &-6 \cos ^4\theta+7 \cos
   ^2\theta+23\bigg)
   \\
   \nonumber
   \\ 
  f^{(4,2,3)}_1&={1\over1440
   \left(\frac{r^2}{a^2}+1\right)^7}\bigg(\frac{20 r^{12} \cos ^2\theta}{a^{12}}-\frac{20 r^{10} \cos ^4\theta}{a^{10}}+\frac{150 r^{10} \cos ^2\theta}{a^{10}}-\frac{1440 r^8 \cos ^6\theta}{a^8}\nonumber\\
   &+\frac{2040
   r^8 \cos ^4\theta}{a^8}-\frac{386 r^8 \cos ^2\theta}{a^8}-\frac{720 r^6 \cos ^6\theta}{a^6}+\frac{1286 r^6 \cos ^4\theta}{a^6}-\frac{419 r^6 \cos ^2\theta}{a^6}\nonumber\\ 
   &-\frac{70 r^4 \cos ^4\theta}{a^4}+\frac{123 r^4 \cos ^2\theta}{a^4}-\frac{18 r^2 \cos ^4\theta}{a^2}+\frac{29 r^2 \cos ^2\theta}{a^2}-\frac{10
   r^{12}}{a^{12}}-\frac{70 r^{10}}{a^{10}}\nonumber\\
   &-\frac{137 r^8}{a^8}-\frac{112 r^6}{a^6}-\frac{44 r^4}{a^4}-\frac{10 r^2}{a^2}-2 \cos ^4\theta+3 \cos ^2\theta-1\bigg)
   \\
   \nonumber\\
   f^{(4,2,3)}_2 &= -{1\over1440 \left(\frac{r^2}{a^2}+1\right)^7}\bigg(-\frac{10 r^{10} \cos ^2\theta}{a^{10}}-\frac{428 r^8 \cos ^4\theta}{a^8}+\frac{368 r^8 \cos ^2\theta}{a^8}\nonumber\\
   &+\frac{720 r^6 \cos ^6\theta}{a^6}-\frac{1034 r^6 \cos
   ^4\theta}{a^6}+\frac{237 r^6 \cos ^2\theta}{a^6}+\frac{138 r^4 \cos ^4\theta}{a^4}-\frac{173 r^4 \cos ^2\theta}{a^4}\nonumber\\ 
   &+\frac{26 r^2 \cos ^4\theta}{a^2}-\frac{35 r^2
   \cos ^2\theta}{a^2}+\frac{10 r^{12}}{a^{12}}+\frac{70 r^{10}}{a^{10}}+\frac{137 r^8}{a^8}+\frac{112 r^6}{a^6}+\frac{44 r^4}{a^4}+\frac{10 r^2}{a^2}\nonumber\\
   &+2 \cos ^4\theta-3 \cos
   ^2\theta+1\bigg)
   \\
   \nonumber\\
   Z_1^{(4,2,3)}&=\frac{Q_1}{ \left(r^2+a^2\cos ^2\theta\right)}-{\tilde b^2\over1440  \left(r^2+a^2\cos ^2\theta\right)\left(\frac{r^2}{a^2}+1\right)^7 }\bigg( \frac{40 r^{10} \cos ^4\theta}{a^{10}}\nonumber\\
   &-\frac{40 r^{10} \cos ^2\theta}{a^{10}}+\frac{240 r^8 \cos ^4\theta}{a^8}-\frac{240 r^8 \cos ^2\theta
   }{a^8}-\frac{412 r^6 \cos ^4\theta}{a^6}+\frac{412 r^6 \cos ^2\theta}{a^6}\nonumber\\
   &+\frac{140 r^4 \cos ^4\theta}{a^4}-\frac{140 r^4 \cos ^2\theta}{a^4}+\frac{36 r^2 \cos
   ^4\theta}{a^2}-\frac{36 r^2 \cos ^2\theta}{a^2}+\frac{10 r^{10}}{a^{10}}+\frac{60 r^8}{a^8}\nonumber\\
   &+\frac{77 r^6}{a^6}+\frac{35 r^4}{a^4}+\frac{9 r^2}{a^2}+4 \cos ^4\theta-4 \cos
   ^2\theta+1\bigg)
\end{align}
Regularity conditions at the supertube locus give the following constraint
\bea
Q_1Q_5 = {Q_5\tilde b^2\over60} + a^2 R_y^2\,,
\eea
and the left and right moving angular momenta are found to be
\bea
J = {Q_5\tilde b^2 \over120 R_y} + {a^2 R_y\over2},\qquad \bar J  ={a^2 R_y\over2}\,.
\eea
\bibliographystyle{JHEP}

\bibliography{VS2}

\end{document}